\documentclass[useAMS,usenatbib]{mn2e}

\usepackage{graphicx}
\usepackage{rotating}
\usepackage{amsmath}
\usepackage{amssymb}
\usepackage{lscape}
\usepackage{natbib}
\usepackage{keyval}
\usepackage{lastpage}
\usepackage{array}
\usepackage{dcolumn}
\usepackage{comment}

\newcommand{\tbgswarm}{\hbox{$T_\mathrm{W}^{\,\mathrm{BC}}$}}

\newcommand{\tbgscold}{\hbox{$T_\mathrm{C}^{\,\mathrm{ISM}}$}}
\newcommand{\fmu}{\hbox{$f_\mu$}}
\newcommand{\fmuOpt}{\hbox{$f_\mu^{Opt}$}}
\newcommand{\fmuIR}{\hbox{$f_\mu^{IR}$}}

\newcommand{\ldust}{\hbox{$L_{\mathrm{d}}$}}

\newcommand{\tauv}{\hbox{$\hat{\tau}_{V}$}}
\newcommand{\tauvISM}{\hbox{$\hat{\tau}_{V}^{ISM}$}}
\newcommand{\ssfr}{\hbox{$\psi_{\mathrm S}$}}
\newcommand{\sfr}{\hbox{$\psi$}}
\newcommand{\mdms}{\hbox{$M_{\mathrm d}/M_\ast$}\,}

\newcommand{\md}{\hbox{$M_{\mathrm d}$}}
\newcommand{\mstar}{\hbox{$M_\ast$}}
\newcommand{\lco}{\hbox{$L^{\prime}_{\rm{CO(1-0)}}$}}
\newcommand{\msun}{\hbox{$\rm{M_{\odot}}$}}
\newcommand{\lir}{\hbox{$L_{\rm{IR}}$}}
\newcommand{\lfir}{\hbox{$L_{\rm{FIR}}$}}

\def\arcsec{\hbox{$^{\prime\prime}$}}

\title[H-ATLAS: Properties of dusty massive galaxies at low and high redshifts]
{{\it Herschel}\thanks{{\it Herschel} is an ESA space observatory with science instruments provided by European-led Principal Investigator consortia and with important participation from NASA.}-ATLAS: Properties of dusty massive galaxies at low and high redshifts.}

\author[K. Rowlands et al.]
{K. Rowlands$^{2,1}$\thanks{E-mail:ker7@st-andrews.ac.uk}, ~L. Dunne$^{3}$, S. Dye$^{2}$, A. Arag\'{o}n-Salamanca$^{2}$, S. Maddox$^{3}$ 
\newauthor E. da Cunha$^{4}$, D.~J.~B. Smith$^{5}$, N. Bourne$^{2,6}$, S. Eales$^{7}$, H.~L. Gomez$^{7}$, I. Smail$^{8}$, 
\newauthor M. Alpaslan$^{9,1}$, C.~J.~R. Clark$^{7}$, S. Driver$^{9,1}$, E. Ibar$^{10}$ R.\,J.~Ivison$^{11,6}$, 
\newauthor A. Robotham$^{9,1}$, M.~W.~L. Smith$^{7}$, E. Valiante$^{7}$
\\ 
   $^{1}$(SUPA) School of Physics \& Astronomy, University of St Andrews, North Haugh, St Andrews, KY16 9SS, UK \\ 
   $^{2}$School of Physics \&\ Astronomy, The University of Nottingham, University Park Campus, Nottingham, NG7 2RD, UK \\  
   $^{3}$Department of Physics and Astronomy, University of Canterbury, Private Bag 4800, Christchurch, New Zealand \\ 
   $^{4}$Max Planck Institute for Astronomy, Konigstuhl 17, 69117, Heidelberg, Germany \\ 
   $^{5}$Centre for Astrophysics, Science \&\ Technology Research Institute, University of Hertfordshire, Hatfield, Herts, AL10 9AB, UK \\ 
   $^{6}$Institute for Astronomy, The University of Edinburgh, Royal Observatory, Blackford Hill, Edinburgh EH9 3HJ, UK \\
   $^{7}$School of Physics \&\ Astronomy, Cardiff University, Queens Buildings, The Parade, Cardiff, CF24 3AA, UK \\ 
   $^{8}$Institute for Computational Cosmology, Department of Physics, Durham University, South Road, Durham DH1 3LE, UK \\ 
   $^{9}$International Centre for Radio Astronomy (ICRAR), University of Western Australia, Crawley, WA6009, Australia \\ 
   $^{10}$Instituto de F\'isica y Astronom\'ia, Universidad de Valpara\'iso, Avda. Gran Breta\~na 1111, Valpara\'iso, Chile \\
   $^{11}$European Southern Observatory, Karl Schwarzschild Strasse 2, D-85748 Garching, Germany \\
}

\begin{document}

\date{}

\pagerange{\pageref{firstpage}--\pageref{lastpage}} \pubyear{2014}

\maketitle

\label{firstpage}

\begin{abstract}
We present a comparison of the physical properties of a rest-frame $250\mu$m selected sample of massive, dusty galaxies from $0<z<5.3$. Our sample comprises 29 high-redshift submillimetre galaxies (SMGs) from the literature, and 843 dusty galaxies at $z<0.5$ from the \emph{Herschel}-ATLAS, selected to have a similar stellar mass to the SMGs. The $z>1$ SMGs have an average SFR of $390^{+80}_{-70}\,$M$_\odot$yr$^{-1}$ which is 120 times that of the low-redshift sample matched in stellar mass to the SMGs (SFR$=3.3\pm{0.2}$\,M$_\odot$yr$^{-1}$). The SMGs harbour a substantial mass of dust ($1.2^{+0.3}_{-0.2}\times{10}^9\,$M$_\odot$), compared to $(1.6\pm0.1)\times{10}^8\,$M$_\odot$ for low-redshift dusty galaxies. At low redshifts the dust luminosity is dominated by the diffuse ISM, whereas a large fraction of the dust luminosity in SMGs originates from star-forming regions. At the same dust mass SMGs are offset towards a higher SFR compared to the low-redshift H-ATLAS galaxies. This is not only due to the higher gas fraction in SMGs but also because they are undergoing a more efficient mode of star formation, which is consistent with their bursty star-formation histories. The offset in SFR between SMGs and low-redshift galaxies is similar to that found in CO studies, suggesting that dust mass is as good a tracer of molecular gas as CO. 
\end{abstract}


\section{Introduction}
\label{Intro}
The first blind submillimetre surveys discovered a population of luminous ($L_\mathrm{IR}>10^{12}$\,L$_\odot$), highly star-forming ($100-1000\,$M$_\odot$yr$^{-1}$), dusty ($10^{8-9}$M$_\odot$) galaxies at high redshift \citep{Smail97, Hughes98, Barger98, Eales99}. These submillimetre galaxies (SMGs) are thought to be undergoing intense, obscured starbursts \citep{Greve05, Alexander05, Tacconi06, Pope08}, which may be driven by gas-rich major mergers \citep[e.g.][]{Tacconi08, Engel10, Wang11, Riechers11, Bothwell13}, or streams of cold gas \citep{Dekel09, Dave10, vandeVoort11a}. Measurements of the stellar masses, star-formation histories (SFHs) and clustering properties of SMGs indicate that they may be the progenitors of massive elliptical galaxies observed in the local Universe \citep{Eales99, Blain02, Dunne03b, Chapman05, Swinbank06, Hainline11, Hickox12}. Due to their extreme far-infrared (FIR) luminosities, it was proposed that SMGs were the high-redshift analogues of local ultra-luminous infrared galaxies (ULIRGs), which are undergoing major mergers. Recent observations \citep{Magnelli12, Targett13} and simulations \citep{Dave10, Hayward11} have suggested that the SMG population is a mix of starbursts and massive star-forming galaxies, with the most luminous SMGs ($L_\mathrm{IR}\sim10^{13}$\,L$_\odot$) being major mergers and lower luminosity SMGs being consistent with turbulent, star-forming disks. There are, however, still considerable uncertainties in the physical properties of SMGs \citep[e.g.][]{Hainline11, Michalowski12}, which affects our view of how SMGs fit into the general picture of galaxy evolution.

SMGs are found to typically reside at $z\sim1-3$ \citep{Chapman05, Chapin09, Lapi11, Wardlow11, Yun11, Michalowski12b, Simpson14}, partly due to the effect of the negative $k$-correction, which allows galaxies which are bright at $>850\mu$m to be detected across a large range in redshift \citep{Blain02}. Due to the long integration times required to survey a large area of sky at $850\mu$m, submillimetre survey volumes at low redshift have until recently been relatively small, leading to difficulties in obtaining a representative sample of dusty galaxies at low redshift. With the launch of the \emph{Herschel Space Observatory} \citep{Pilbratt10}, we can now get an unprecedented view of dust in local galaxies. \emph{Herschel} observed at FIR--submillimetre wavelengths across and beyond the peak of the dust emission, making it an unbiased tracer of the dust mass in galaxies. The \emph{Herschel} Astrophysical TeraHertz Large Area Survey (H-ATLAS, \citealt{Eales_ATLAS10}) is the largest area extra-galactic survey carried out with \emph{Herschel} and has allowed us to quantify the amount of dust in galaxies at low redshift. By studying galaxies selected at $250\mu$m, \citet{Smith12} found an average dust mass of $9.1\times10^{7}$\,M$_{\odot}$ in local ($z<0.35$) dusty galaxies. Furthermore, the dust mass in galaxies is found to increase by a factor of $3-4$ between $0<z<0.3$ \citep{Dunne11, Bourne12a}, which may be linked to higher gas fractions in galaxies at earlier epochs \citep{Geach11, Tacconi13, Combes13}.

The question of how the modes of star formation in SMGs relates to those in local star-forming galaxies warrants a comparison between galaxy samples. Comparisons between SMGs and the low redshift galaxy population has been carried out for small galaxy samples, e.g. \citet{Santini10} compared the properties of 21 SMGs to 26 local spirals from SINGS \citep{Kennicutt03} and 24 local ULIRGs from \citet{Clements10} and found that SMGs have dust-to-stellar mass ratios 30 times larger than local spirals, and a factor of 6 more than local ULIRGs. However, a comparison to large representative samples of the general dusty galaxy population has not yet been carried out. In this paper we investigate the physical properties of dusty galaxies over a wide range in cosmic time, utilising carefully selected samples of high and low redshift galaxies which occupy comparable co-moving volumes of $\sim 10^{8}$\,Mpc$^{3}$. 

We describe our sample selection in \S\ref{sec:sample_selection} and spectral energy distribution (SED) fitting method to explore the properties of SMGs in \S\ref{sec:SED_fitting}. Our results are presented in \S\ref{sec:results} and our conclusions are in \S\ref{sec:conclusions}. We adopt a cosmology with $\Omega_m=0.27,\,\Omega_{\Lambda}=0.73$ and $H_o=71\, \rm{km\,s^{-1}\,Mpc^{-1}}$.

\section{Sample selection}
\label{sec:sample_selection}
In order to investigate the physical properties of dusty galaxies over a range of redshifts, we construct a sample selected at $\sim250\mu$m rest-frame wavelength. This comprises panchromatic photometry of low redshift galaxies from the H-ATLAS Phase 1 catalogue, and a sample of high-redshift SMGs presented in \citet{Magnelli12}.

\subsection{Low redshift H-ATLAS sample}
The H-ATLAS is a $\sim$590 deg$^2$
survey undertaken by \emph{Herschel}
at 100, 160, 250, 350 and 500$\mu$m to provide an
unbiased view of the submillimetre Universe. Observations were carried
out in parallel mode using the PACS \citep{PACS10} and SPIRE
\citep{Griffin10} instruments simultaneously. The observations in the Phase 1 field cover an area of $\sim$161 deg$^2$
centred on the Galaxy And Mass Assembly (GAMA) 9, 12 and 15 hr equatorial fields \citep{GAMA_Driver11}. Details of the map making can be found in \citet{Pascale11}, \citet{Ibar10} and Smith et al. (in prep). We use the catalogue of $\geq$5$\sigma$ detections in the 250$\mu$m band \citep[][Valiante et al. in prep.]{Rigby11} produced using the MAD-X algorithm (Maddox et al. in prep). Fluxes at 350 and 500$\mu$m are measured at the location of the 250$\mu$m fitted position. A likelihood-ratio analysis \citep{SuthSaund92, Smith11a} was then performed to match the 250$\mu$m sources to SDSS DR7 \citep{SDSS_DR7} galaxies with $r<22.4$. This method accounts for the possibility that the true counterpart is below the optical magnitude limit and uses the positional uncertainty as well as empirical magnitude priors to estimate the probability (reliability) of a submillimetre source being the true association of a given optical counterpart. SDSS sources with reliability $R\geq0.8$ are considered to be likely matches to submillimetre sources.

PACS 100 and $160\mu$m flux densities were measured for all 250$\mu$m sources \footnote{Except those with SDSS $r$-band isophotal major axis (isoA) $>30$\arcsec{}, where reliable PACS fluxes cannot be obtained due to aggressive high-pass filtering in the maps. This issue will be rectified in the public data release.} by placing apertures at the SPIRE positions. Aperture photometry for extended SPIRE sources was also performed according to the procedure described in \citet{Rigby11}. The final catalogue has 103721 sources detected at 250$\mu$m at $\geq 5\sigma$, with flux estimates in each of the other four bands at that position. The $5\sigma$ noise levels were 130, 130, 30, 37 and 41mJy per beam at 100, 160, 250, 350 and 500$\mu$m, respectively; with beam sizes of $9$, $13$, 18, 25 and 35\arcsec{} in these bands.

From this catalogue, there are 29787 reliable optical counterparts to H-ATLAS sources; with 14920 sources having good quality spectroscopic redshifts, and 14867 sources having photometric redshifts. The contamination rate by false identifications is given by $\sum(1-R)$ following \citet{Smith11a}, and is expected to be 3.8 per cent. The median and 84th--16th percentile range of $250\mu$m flux densities of sources with reliable counterparts with good quality spectroscopic redshifts at $z<0.5$ is $0.05^{+0.04}_{-0.01}$\,Jy. Around two-thirds of the sources without reliable optical counterparts are unidentified because their counterparts lie below the optical magnitude limit, and these sources mostly reside at $z>0.5$ (see \citealt{Dunne11}). The remaining unidentified sources are believed to have a counterpart in the SDSS catalogue but we are unable to unambiguously identify the correct counterpart in these cases due to near neighbours and the non-negligible probability of a background galaxy of the same magnitude being found at this distance. The optically identified sources are believed to be a representative sample of all H-ATLAS sources at $z\leq0.35$ \citep{Smith12}.

The optically identified counterparts were combined with GAMA data \citep{GAMA_Driver11, GAMA_Robotham10, GAMA_Baldry10} to provide $r$-band defined
matched aperture photometry as described in \citet{GAMA_Hill11}. The $FUV$ and $NUV$ photometry is from \emph{GALEX} \citep[][Seibert et al. in prep.]{GALEX_Martin05, GALEXDR3}, and is a reconstruction of the true UV flux of a given GAMA object. This accounts for cases where multiple GAMA and GALEX objects are associated with each other. Optical $ugriz$ magnitudes are derived from SDSS DR6 imaging \citep{AM09} and near-infrared $YJHK$ photometry are from UKIDSS-LAS imaging \citep{UKIDSS-LAS}. All UV-NIR photometry has been galactic extinction corrected. Spectroscopic redshifts are included from the GAMA, SDSS and 6dFGS catalogues for 14490 sources at $z<0.5$; where spectroscopic redshifts are not available we use ANN$z$ \citep{ANNz04} neural network photometric redshifts from \citet{Smith11a}. \citet{Smith11a} estimate the completeness of the H-ATLAS sample as a function of redshift by calculating the total number sources that we would expect to have a counterpart above the SDSS magnitude limit in H-ATLAS; we refer the reader to \citet{Smith11a, Dunne11} for further details. 

\subsection{High-redshift SMG sample}
\label{sec:High_redshift_sample}
Estimates from submillimetre photometric redshift studies suggest that $\sim 50$ percent of H-ATLAS sources are at $z>1$ \citep{Lapi11, Pearson13}, however, identifications to these submillimetre sources are not currently available due to the relatively shallow ancillary multiwavelength data. We therefore rely on publicly available measurements of high redshift submillimetre-selected galaxies (SMGs) with robust optical counterparts and spectroscopic redshifts in the literature. We utilise the compilation of SMGs in \citet[][hereafter M12]{Magnelli12} taken from blank field (sub)millimetre surveys ($850-1200\mu$m) which have robust counterparts identified with deep radio, interferometric submillimetre and/or mid-infrared (MIR) imaging from \citet{Chapman05, Pope2006, Bertoldi07, Ivison07, Younger07, Pope08, Chapin09, Younger09, Coppin10, Biggs11}, Aravena et al. in prep. 
The spectroscopic redshifts in the M12 sample are from \citet{Borys04, Chapman05, Pope08, Daddi09a, Coppin10}, Danielson et al. in prep, Capak et al. in prep. 
The SMGs are located in fields which have excellent multiwavelength coverage (GOODS-N, ECDFS, COSMOS and Lockman Hole), which is required in order to derive statistical constraints on galaxy physical properties using SED fitting. Most crucially, all of the galaxies in our sample have well sampled coverage of the peak of the dust emission in the FIR, which allows us to derive robust constraints on the dust luminosity of our SMGs. This coverage of the dust peak is not available for all sources in larger samples of SMGs (e.g. those from \citealp{Chapman05}). Estimates of the dust luminosity and temperature are therefore often subject to assumptions about the SED shape with constraints based on only one or two FIR--submillimetre measurements. 

In M12 the radio or MIPS counterparts to the SMGs were matched within 3\arcsec{} to \emph{Spitzer} Multiband Imaging Photometer (MIPS; \citealt{Rieke04MIPS}) $24\mu$m positions associated with PACS and SPIRE data at $70\mu$m, $100\mu$m, $160\mu$m, $250\mu$m, $350\mu$m and $500\mu$m from the PACS Evolutionary Probe (PEP; \citealt{Lutz11}) and \emph{Herschel} Multi-tiered Extragalactic Survey (HerMES; \citealt{Oliver12}). The reduction of the HerMES maps is described in \citet{Smith_HerMES12}, and cross-identifications of $24\mu$m and SPIRE sources were performed in \citet{Roseboom10}. The PACS and SPIRE fluxes of the sources were extracted by fitting a point spread function (PSF) at the $24\mu$m position, which allows the flux of blended FIR sources to be recovered. Additionally, the inherent association of a SPIRE source with a more accurate $24\mu$m position allows for relatively easy identification of multiwavelength counterparts. 

M12 present photometry for 61 galaxies, however, we only consider the 46 SMGs which are unlensed. This is because M12 found difficulty in obtaining good quality optical-NIR photometry which is required for deriving constraints on stellar masses and SFRs. We also conservatively exclude 6/46 sources listed in M12 which have multiple robust counterparts to the submillimetre source where both counterparts are at the same redshift. These systems are thought to be interacting, so the submillimetre emission is thought to originate from both sources and there is no way to quantify the individual contribution of each counterpart to the submillimetre emission. We note that other sources in our sample with single robust counterparts may also be interacting systems, this is discussed in \S\ref{sec:multiple_sources}. Four sources (LESS10, LOCK850.03, LOCK850.04 and LOCK850.15) have multiple robust counterparts for which only one counterpart has a spectroscopic redshift. Following M12 we include these galaxies in our sample, as the $24\mu$m and radio flux densities of the spectroscopic counterpart agree with the infrared luminosity computed from the FIR-submillimetre flux densities. This supports the assumption that the submillimetre emission originates from one counterpart. The inclusion or exclusion of these galaxies does not change our results. We include four galaxies which have a $<3\sigma$ detection above the confusion limit in at least one of the SPIRE bands so we do not bias our sample towards sources with warm dust temperatures. M12 note that one of these (GN15) is isolated and so its measured flux densities should be reliable, however for the other three (GN5, GN20, GN20.2) the FIR emission is confused with that from near neighbours, which may lead to some overestimation of their FIR fluxes. For GN20 we use the \emph{Herschel} photometry from \citet{Magdis11}, which has been carefully deblended based on $24\mu$m and radio positional priors. We use different symbols for these confused sources in later figures so that any systematic biases relative to the rest of the sample can be easily seen.

We match the counterpart positions presented in M12 to ancillary optical--MIR data using a 1\arcsec{} search radius for optical data\footnote{1\arcsec{} corresponds to 8.5\,kpc at $z=2$ for our adopted cosmology.} and a 2\arcsec{} search radius for \emph{Spitzer} Infrared Array Camera (IRAC; \citealt{Fazio04}) data. We only include a galaxy counterpart in our sample if it has IRAC data, as we expect $24\mu$m detected galaxies to also have IRAC data. Across all fields we find that six sources which were included in M12 do not have optical matches within 1\arcsec{}.
In the COSMOS field we use the broad, medium and narrow band photometry as presented in \citet{Ilbert09} and \citet{Salvato09}. The public \emph{Spitzer} IRAC photometry was retrieved from the COSMOS archive\footnote{http://irsa.ipac.caltech.edu/data/COSMOS/}. The GOODS-N multiwavelength catalogue is briefly described in \citet{Berta10,Berta11} and includes PSF-matched photometry from \emph{HST} ACS $bviz$ (version 1.0), FLAMINGOS $JHK$\footnote{The KPNO 4m FLAMINGOS data were kindly provided by Mark Dickinson, Kyoungsoo Lee and the GOODS team.} and IRAC 3.6, 4.5, 5.8, 8.0$\mu$m obtained with the ConvPhot code \citep{Grazian06}, spectroscopic redshifts from \citet{Barger08} and associated \emph{GALEX}, $U$-band, radio and X-ray fluxes. Deep CFHT WirCAM $K_s$ band photometry was taken from \citet{Wang10} and 24 and $70\mu$m MIPS data are from \citet{Magnelli11}. In ECDFS we use the compilation of photometry for SMGs presented in \citet{Wardlow11} from the MUSYC \citep{MUSYC,Taylor09}, IRAC photometry from SIMPLE \citep{Damen11} and GOODS/VIMOS $U$-band data from \citet{Nonino09}. In the Lockman Hole we use the photometry described in \citet{Fotopoulou12}, which comprises UV data from \emph{GALEX}, Large Binocular Telescope ($U, B, V, Y, z'$) and Subaru ($R_c, I_c, z'$) photometry, $J$ and $K$ photometry from the UKIRT and MIR data from IRAC. We follow the recommendations in each catalogue and apply the relevant offsets to correct all of the photometry to total magnitudes. Additionally, we have removed any spurious or problematic photometry, in particular COSMOS medium band photometry where we suspect that strong nebular emission lines contribute significantly to the flux. Deboosted millimetre photometry is provided for some sources in M12 where available from \citet{Greve04, Bertoldi07, Greve08, Perera08, Scott08, Chapin09, Austermann10, Scott10}. The final sample comprises 34 SMGs with robust counterparts and panchromatic data from the rest-frame UV to the submillimetre\footnote{The photometry for the SMGs are available electronically from VizieR: http://vizier.u-strasbg.fr/viz-bin/VizieR.}.

In order to account for additional uncertainties, for example, in deriving total flux measurements and photometric calibration for the wide array of multiwavelength data, we add in quadrature a calibration error to the catalogue photometric errors. For optical, near-infrared (NIR), MIR and FIR bands we add in quadrature 20 per cent of the flux. We add 30 per cent of the flux for (sub)millimetre ($\geq850\mu$m) data to account for calibration errors, the uncertainty in deboosting the fluxes and source blending. For sources which are not detected, we set the fluxes to upper limits as detailed in the respective catalogues; these are typically $5\sigma$ upper limits in the optical-NIR bands and $3\sigma$ upper limits longwards of $24\mu$m.

\subsubsection{SMG multiplicity}
\label{sec:multiple_sources}
Another source of uncertainty in our SMG sample is source multiplicity. Using ALMA data, \citet{Hodge13} estimated that $35-50\%$ of single dish-detected SMGs are comprised of multiple sources. The fraction of SMGs which are multiple is likely to be slightly lower in our sample, as we have removed SMGs which have more than one robust counterpart at the same redshift. The seven SMGs in ECDFS observed by \citet{Hodge13} confirm that 4/7 SMGs are single sources (LESS10, LESS11, LESS17, LESS18), with the ALMA position in good agreement with the radio position given in M12. One source (LESS40) was not detected above the $3\sigma$ limit; this maybe due to the relatively low quality image of this source, although it is possible that this source is comprised of multiple faint submillimetre sources which are below the ALMA detection threshold. The other two sources (LESS67 and LESS79) are comprised of multiple submillimetre counterparts, with only one component coincident with the position of the robust radio/MIPS counterpart identified in \citet{Biggs11}. Since the FIR photometry for the multi-counterpart SMGs are blended, it is possible that the dust masses are overestimated. The flux for the ALMA component coincident with the LABOCA robust counterpart position amounts to 0.73 and 0.25 of the total ALMA flux in these two blended cases. Given that the flux of the ALMA counterparts are 0.87 and 0.43 of the deboosted LABOCA flux for LESS67 and LESS79, respectively, we may expect that the dust masses would be overestimated by similar factors. The change in dust mass for LESS67 is within the $1\sigma$ uncertainty on the dust masses from the SED fitting, and the change in dust mass for LESS79 is within  the $3\sigma$ uncertainty on the dust mass. 
We find that the dust masses for these blended sources using the LABOCA fluxes are not outliers in our sample of SMGs, and are similar to the dust masses of SMGs confirmed to have a single counterpart; therefore blending does not affect our conclusions.

\section{SED fitting}
\label{sec:SED_fitting}
The wealth of multiwavelength coverage for our sample of dusty galaxies allows us to derive physical properties using SED fitting techniques. Due to a lack of FIR data, studies of SMGs have often derived dust luminosities and SFRs based upon fitting SEDs to $850\mu$m photometry alone. The availability of \emph{Herschel} data across the peak of the dust emission provides better constraints on the dust luminosity than previous studies \citep[e.g.][]{Chapman05}, see M12 for a review.

We use a modified version of the physically motivated method of \citet*[][hereafter DCE08\footnote{The \citet*{DCE08} models are publicly available as a user-friendly model package {\sc magphys} at www.iap.fr/magphys/.}]{DCE08} to recover the physical properties of the galaxies in our sample.
In this method the UV-optical radiation emitted by stellar populations is
absorbed by dust, and this absorbed energy is matched to that re-radiated in the FIR. 
Spectral libraries of 50000 optical models with stochastic SFHs, and 50000
infrared models, are produced at the redshift of each galaxy in our
sample, containing model parameters and synthetic photometry from the
UV to the millimetre. The model libraries are constructed from parameters which have prior distributions designed to reproduce the range of properties found in galaxies. The optical libraries are produced using the spectral evolution of stellar populations calculated from the latest version of the population synthesis code of \citet{BC03}. The stellar population models include a revised prescription for thermally-pulsing asymptotic giant branch (TP-AGB)
stars from \citet{Marigo_Girardi07}. A \citet{Chabrier03} Galactic-disk Initial Mass Function (IMF) is assumed. The libraries contain model spectra with a wide range of SFHs, metallicities and dust attenuations. The two-component dust model of \citet{CF00} is used to calculate the attenuation of
starlight by dust, which accounts for the increased attenuation of
stars in birth clouds compared to old stars in the ambient interstellar medium (ISM).
The model assumes angle-averaged spectral properties and so does not include 
any spatial or dynamical information. Hayward \& Smith (in prep) find that physical properties derived using {\sc magphys} are robust to projection effects associated with different viewing angles.

The infrared libraries contain SEDs comprised of four different temperature dust components, from which the dust mass (${M}_\mathrm{d}$) is calculated. In stellar birth clouds, these components are polycyclic aromatic hydrocarbons (PAHs), hot dust (stochastically heated small grains with a temperature $130-250$\,K), and warm dust in thermal equilibrium ($30-60$\,K). In the diffuse ISM the relative fractions of these three dust components are fixed, but an additional cold dust component with an adjustable temperature between 15 and 25\,K is added. The dust mass absorption coefficient $\kappa_{\lambda} \propto \lambda^{-\beta}$ has a normalisation of $\kappa_{850}=0.077\,\rm{m}^2 \,\rm{kg}^{-1}$ \citep{SLUGS00a}. A dust emissivity index of $\beta=1.5$ is assumed for warm dust, and $\beta=2.0$ for cold dust, following studies which support a value of $\beta$ dependant on the temperature of the dust components \citep{DH02, MWLSmith12, Davies13}, see also the review in \citet{DE01}. The prior distributions for the temperature of warm dust in birth clouds (\tbgswarm), and the temperature of cold dust in the diffuse ISM (\tbgscold) are flat (see Fig.~\ref{fig:priors}), so that all temperatures within the bounds of the prior have equal probability in the model libraries.

The attenuated stellar emission and dust emission models in the two spectral libraries are combined using a simple energy balance argument, that the energy absorbed by dust in stellar birth clouds and the diffuse ISM is re-emitted in the FIR. In practise, this means that each model in the optical library is matched to models in the infrared library which have the same fraction of total dust luminosity contributed by the diffuse ISM (\fmu), within a tolerance of 0.15, and are scaled to the total dust luminosity\footnote{Integrated between 3 and $1000\mu$m.} \ldust. Statistical constraints on the various parameters of the model are derived using the Bayesian approach described in DCE08. Each observed galaxy SED is compared to a library of stochastic models which encompasses all plausible parameter combinations. For each galaxy, the marginalised likelihood distribution of any physical parameter is built by evaluating how well each model in the library can account for the observed properties of the galaxy (by computing the $\chi^{2}$ goodness of fit). This method ensures that possible degeneracies between model parameters are included in the final probability density function (PDF) of each parameter. The effects of individual wavebands on the derived parameters are explored in DCE08, and \citet{Smith12}, but we emphasise the importance of using the \emph{Herschel} FIR-submillimetre data to sample the peak of the dust emission and the Rayleigh-Jeans slope in order to get reliable constraints on the dust mass and luminosity \citep{Smith13}.

The {\sc magphys} code is modified from the public version to take into account flux density upper limits in the $\chi^2$ calculation to give additional constraints on physical parameters. If the flux upper limit is above the model SED, the upper limit does not contribute to the $\chi^2$ value. When the model SED violates the flux upper limit, the flux upper limit is treated like all the other detected photometry by including the upper limit as a flux density (with associated photometric error) in the $\chi^2$ calculation. Additionally, we modify the priors to take into account areas of parameter space which are not explored with the standard {\sc magphys} libraries. This is important when studying a wide variety of galaxies from quiescent systems to highly obscured starburst galaxies. \S\ref{sec:Model_priors} and Appendix \ref{sec:Comparison of priors} outline the standard priors which are more applicable to low redshift galaxies, and also describes the modified priors which better suit the high-redshift SMGs.

\begin{figure*}
\centering
\includegraphics[trim=0mm 0mm 0mm 1mm, clip=true, width=1.0\textwidth]{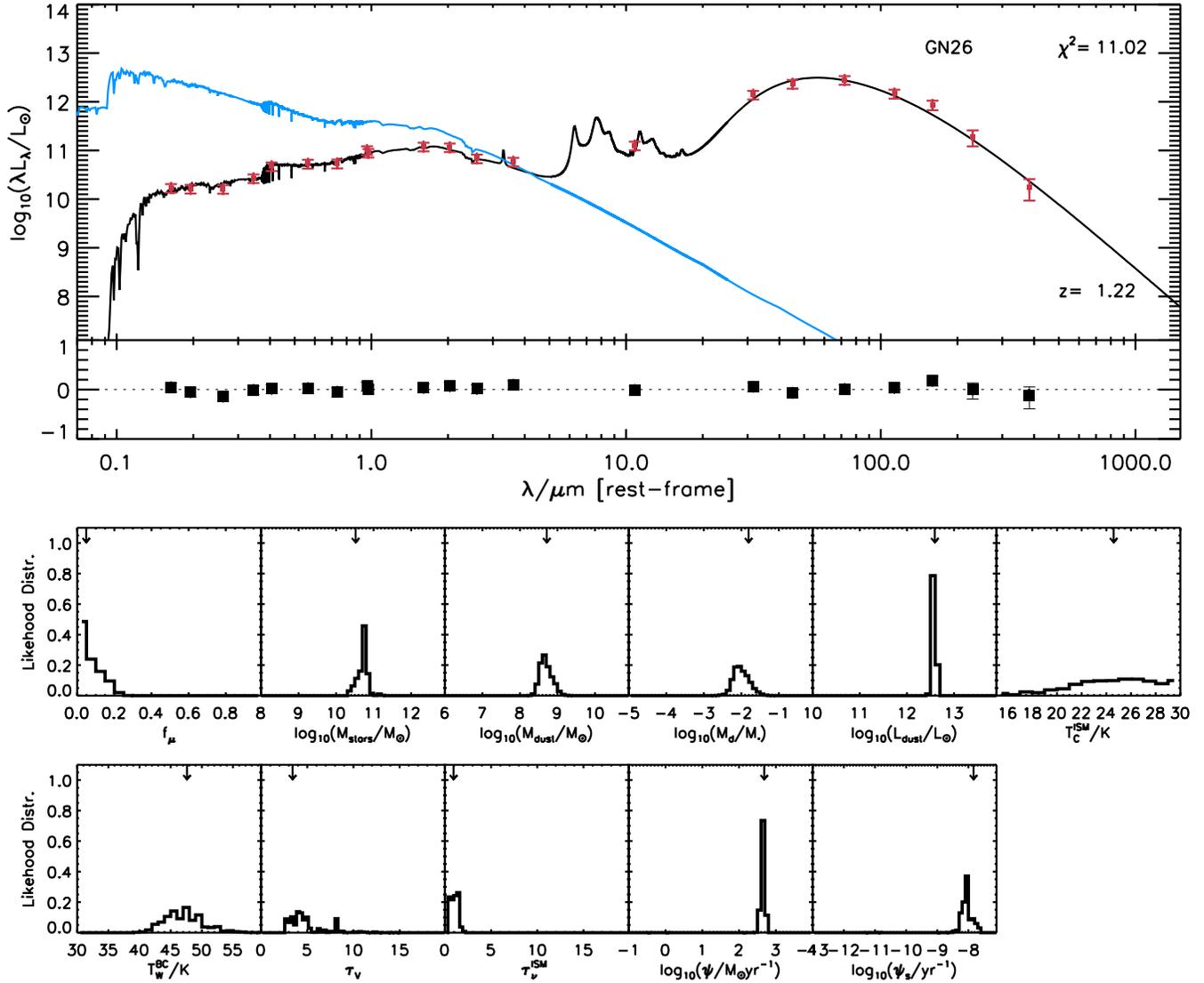}
\caption[Example best-fit rest-frame SED of a high-redshift submillimetre galaxy]{{\bf Top}: Example best-fit rest-frame SED of a high-redshift submillimetre galaxy, with observed photometry (red points) from the rest-frame UV to the submillimetre. Errors on the photometry are described in \S\ref{sec:High_redshift_sample}. The black line is the best-fit model SED and the blue line is the unattenuated optical model. {\bf Bottom}: Probability density functions (PDFs) for each physical parameter are shown for this submillimetre galaxy, with the best-fit model values shown as arrows above each parameter PDF. The parameters are (from left to right): \fmu, the fraction of total dust luminosity contributed by the diffuse ISM; ${M}_\ast/$M$_\odot$, stellar mass; ${M}_\mathrm{d}/$M$_\odot$, dust mass; \mdms, dust-to-stellar mass ratio; \ldust/L$_\odot$, dust luminosity; \tbgscold/K, temperature of the cold diffuse ISM dust component; \tbgswarm/K, temperature of the warm dust component in birth clouds; \tauv, total effective $V$-band optical depth seen by stars in birth clouds; \tauvISM, effective $V$-band optical depth in the ambient ISM; \sfr/M$_\odot$yr$^{-1}$, the star-formation rate (SFR); and \ssfr/yr$^{-1}$, the specific star-formation rate (SSFR). The SSFR and SFR are averaged over the last $10^7$ years, although in this example the result is insensitive to changes in the timescale over which the SFR is averaged.}
\label{fig:SED_example}
\end{figure*}

An example best-fit SED and set of PDFs are shown in Fig.~\ref{fig:SED_example}. The parameters of interest are 
\fmu, the fraction of total dust luminosity contributed by the diffuse ISM;
${M}_\ast/$M$_\odot$, stellar mass;
${M}_\mathrm{d}/$M$_\odot$, dust mass;
\mdms, dust-to-stellar mass ratio;
\ldust/L$_\odot$, dust luminosity;
\tauv, total effective $V$-band optical depth seen by stars in birth clouds;
$\tauvISM$, the effective $V$-band optical depth in the ambient ISM;
\sfr/M$_\odot$\,yr$^{-1}$, the SFR; and 
\ssfr/yr$^{-1}$, specific star-formation rate (SSFR).
For more details of the method we refer the reader to DCE08.

\subsection{Model Priors}
\label{sec:Model_priors}
The `standard' priors which are appropriate for low redshift galaxies are described in detail in DCE08 and were also used in \citet{Smith12} to derive the properties of low redshift H-ATLAS galaxies similar to those in this work. Initial tests with the standard priors showed that there were very few models which had a high enough SSFR to provide a good fit to the photometry of all of the high redshift SMGs. We created modified priors to accommodate a wider range of galaxy characteristics, allowing for higher dust attenuation and SSFR than observed in most low-redshift galaxies. It is not clear whether all SMGs are similar to local ULIRGs with an obscured central starburst, as many show evidence for more extended star formation \citep[e.g.][]{Tacconi08, Hainline09, Swinbank11, Targett13}. Our modified priors (henceforth called `SMG priors') are a hybrid between the ULIRG priors described in \citet{dC10b} and the standard model libraries. A summary of the differences in the prior distributions and how the choice of priors affects our results is given in Appendix~\ref{sec:Comparison of priors}.

\section{Physical properties of rest-frame $~250\mu$\lowercase{m} selected galaxies}
\label{sec:results}
The best-fit SEDs of the 34 SMGs are shown in Fig.~\ref{fig:SEDs}. Evidence from X-ray studies suggest that many SMGs host an AGN \citep{Alexander05}. Indeed some SMGs in our sample show excess emission in the rest-frame NIR, which may be due to dust heated to high temperatures by an obscured AGN \citep{Hainline11}. The {\sc magphys} SED models do not include a prescription for AGN emission and so we must assess the impact that AGN emission may have on the parameters. The details of this process and the results are discussed in Appendix~\ref{sec:AGN} but in brief, we select galaxies at z>1 with power law emission in the NIR from the $S_{24}/S_{8.0}-S_{8.0}/S_{4.5}$ diagram from \citet{Ivison04}, with $S_{8.0}/S_{4.5}>1.65$ \citep{Coppin10}. We find 6/34 galaxies are classed as AGN in this way (AzLOCK.01, AzLOCK.10, AzTECJ100019+023206, LOCK850.04, LOCK850.15 and GN20\footnote{Although the observed $S_{8.0}/S_{4.5}$ colour traces the rest-frame $1.6\mu$m stellar bump at $z\sim4$, we retain GN20 in our AGN sample as \citet{Riechers13} found that GN20 has an obscured AGN from power-law emission in the rest-frame MIR spectrum.}). Following the method of \citet{Hainline11}, we subtract a power-law with $f_{\lambda} \propto \lambda^{\alpha}$, where $\alpha=2$ or 3, from all photometry shortwards of $8\mu$m (observed), incrementally adjusting the power-law contribution at $8\mu$m to achieve the best-fit.

In the following results we use the best-fit power-law subtracted values for the four AGN with weak power law components (AzLOCK.1, AzLOCK.10, LOCK850.15 and GN20). We use the results derived using a power-law slope of $\alpha=3$ as this provides the best-fit to the data, except for the case of GN20 where the data is best-fit by a power-law slope of $\alpha=2$. We exclude AzTECJ100019+023206 as the lack of reliable photometry makes the AGN power-law fraction difficult to constrain, and LOCK850.04 is excluded as the uncertainties on the parameters due to subtraction of the dominant power-law  are too large to make this galaxy a useful member of the sample. The subtraction of a power-law from the photometry of the AGN hosts in all cases results in a better SED fit indicated by a lower $\chi^2$. The galaxies with power law emission comprise a small minority of the SMG sample, and the choice of whether to subtract the power law or not, or exclude them from the sample, does not change our conclusions. 

After subtracting the best-fitting power-law slope from the optical-MIR photometry, as expected we typically see a decrease in the stellar mass of the AGN dominated SMGs.
However, an increase in the stellar mass occurs in some cases because the optical depth increases (albeit with rather large uncertainty). The stellar mass changes by slightly more than the error represented by the 84th--16th percentile range on each individual galaxy PDF (on average $\pm0.11$\,dex). We find that the median-likelihood \fmu\,, SFR, SSFR, and \tauv\, move slightly but are typically within the error represented by the 84th--16th percentile range on each individual galaxy PDF.

We exclude LOCK850.17 and LESS17 from our final sample because there is a large discrepancy between the photometric and spectroscopic redshift. This was also noted for LOCK850.17 in \citet{Dye08}, who propose that the spectroscopic redshift is from a background source blended with a foreground galaxy which dominates the flux measurements. Furthermore, \citet{Simpson14} found that LESS17 has a photometric redshift of $1.51^{+0.10}_{-0.07}$. We exclude GN20.2 as the low signal-to-noise optical-NIR photometry does not allow us to obtain reliable constraints on the physical parameters for this source. The final sample comprises 29 SMGs with $0.48<z<5.31$.

To create a low redshift comparison sample, we fit the UV--millimetre SEDs of 18869 low redshift ($0.005<z<0.5$) H-ATLAS galaxies using a similar method to \citet{Smith12}. These sources are selected to have a reliability $>0.8$ of being associated with an optical counterpart in the SDSS $r$-band catalogue and have multiwavelength photometry from the GAMA survey \citep[see][]{Smith12}. Updated PACS and SPIRE fluxes in all bands are utilised even if they are low signal-to-noise, as this provides more constraint on the SED than setting undetected fluxes to upper limits \citep{Smith13}. To ensure that we only include galaxies which have good photometry, we reject 3856 galaxies which have a less than 1 per cent chance that their photometry is well described by the best-fit model SED, see \citealt{Smith12} for details. Galaxies which are excluded from the sample have problems with AGN contamination or issues with photometry. This can happen where the optical photometry is not equivalent to `total light' if the SExtractor source detection used by GAMA \citep{GAMA_Hill11} deblended single objects, or had stellar contamination, for example. Given the wide parameter space of the {\sc magphys} libraries, galaxies with physically plausible SEDs should be well-fit by our models.

In this study we use the 15013 galaxies at $0.005<z<0.5$ whose photometry is well described by the best-fit model SED. We make two comparisons: one between all the low-redshift H-ATLAS galaxies and the SMGs in order to study the diversity of galaxies which are selected at approximately rest-frame $\sim 250\mu$m and secondly between a stellar mass-matched sample at high and low redshift, in order to determine how the properties of massive submillimetre selected galaxies differ over cosmic time. 

\begin{figure}
\centering
\includegraphics[scale=0.85]{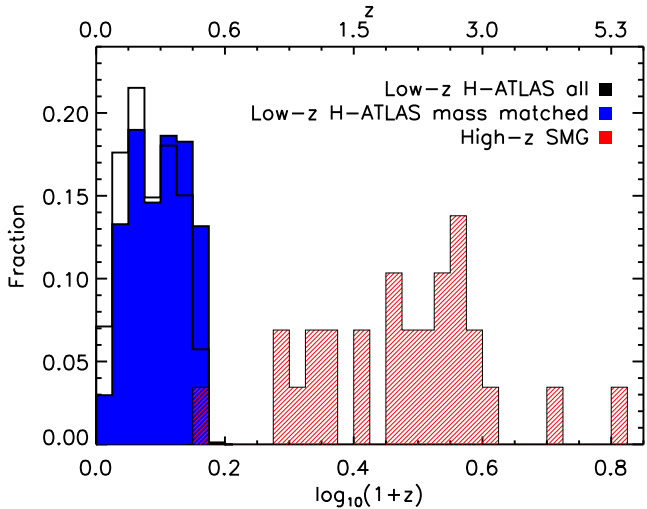}
\caption[Redshift distributions of the low redshift H-ATLAS mass-matched sample and the SMGs.]{Redshift distribution of the entire low redshift H-ATLAS sample (black open histogram), low redshift mass-matched sample (blue solid histogram) and the SMGs (red hatched histogram). The redshift distribution of the 29 SMGs has a mean of $\overline{z}=2.13$, the mean redshift of the 843 low redshift galaxies with a similar stellar mass to the high-redshift sample is $\overline{z}=0.26$, which is similar to that of the entire $z<0.5$ H-ATLAS sample ($\overline{z}=0.22$).}
\label{fig:z_dist}
\end{figure}

To construct the stellar mass-matched sample we split the SMG sample into median-likelihood stellar mass bins of 0.2\,dex width and randomly picked galaxies in the same stellar mass bin from the H-ATLAS sample, such that both distributions matched. We pick the maximum number of H-ATLAS galaxies such that we can still approximately match the SMG stellar mass distribution (30 times the number of SMGs). Even so, there is a lack of H-ATLAS galaxies with the very highest stellar masses with 23/30 (77\%) of galaxies missing from the highest stellar mass bin centred on $10^{11.7}\,$M$_\odot$. Of the total low redshift stellar mass-matched sample (843 galaxies) only $3\%$ of galaxies are missing from the mass-matched sample. The final $\sim250\mu$m rest-frame selected sample comprises 29 SMGs ($\overline{z}=2.13$) and 843 low redshift galaxies from H-ATLAS ($\overline{z}=0.26$) of a similar stellar mass to the high-redshift sample. The redshift distributions of the samples are shown in Fig.~\ref{fig:z_dist}.

The samples investigated in this paper are not typical of the general galaxy population, but represent the most infrared-luminous galaxies at their respective redshifts. We note that the high-redshift SMG sample is not intended to be evolutionarily linked to the low redshift H-ATLAS galaxies. SMGs are likely to rapidly exhaust their gas supply within a few hundred Myr, and are unlikely to be dusty enough to be detected in the H-ATLAS sample at low redshift. The low redshift descendants of the SMGs are thought to be massive ellipticals \citep[e.g.][]{Eales99,Simpson14}. We may glimpse a transitional period where once high redshift dusty starbursts are transitioning onto the red sequence and yet still retain some ISM. Such dusty early-type galaxies have been observed in the H-ATLAS sample by \citet{Rowlands12}, and have comparable stellar masses to the SMG sample.

The selection effects in the high-redshift sample (e.g. the need for radio counterparts, requirement of spectroscopic redshifts, and panchromatic SED coverage) are rather complex, which can result in a biased sample of SMGs. M12 have examined the selection effects in detail and conclude that high infrared luminosity ($L_{\rm{IR}}\geq10^{12.5}$L$_\odot$) SMGs with spectroscopic redshifts are representative of the parent SMG population, and the high luminosity star-forming galaxy population in general. In \S\ref{Sample_params} we therefore concentrate our analysis on these FIR luminous SMGs, which are typically at $z>1$. At lower FIR luminosities the SMG sample shifts to lower redshift galaxies with cooler temperatures and less extreme properties, which produces some overlap between the SMG and low redshift samples. Quantitative comparisons between the high and low redshift samples should therefore be interpreted within the selection functions of the samples.

\begin{figure*}
\centering
\includegraphics[width=1.0\textwidth]{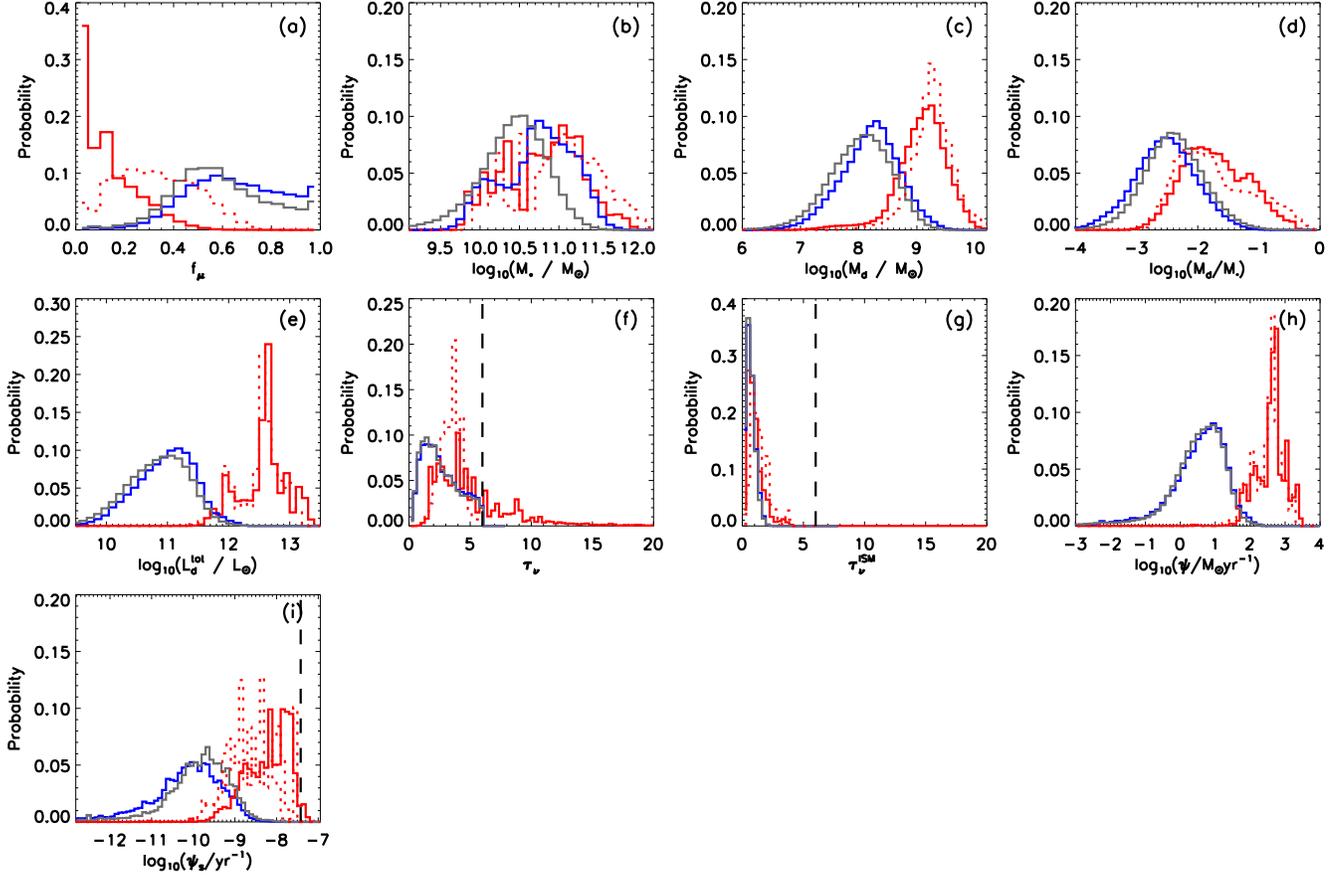}
\caption[Mean probability density functions of the low redshift and high-redshift samples.]{Mean probability density functions (PDFs) of the entire $z<0.5$ H-ATLAS sample (grey solid line), the low redshift mass-matched sample (blue solid line) and high redshift $z>1$ SMGs (red solid line). We also show the mean PDF for the high-redshift SMG sample using the standard priors (red dotted line). The parameters are (from left to right): \fmu, the fraction of total dust luminosity contributed by the diffuse ISM; ${M}_\ast/\rm{M}_\odot$, stellar mass; ${M}_\mathrm{d}/\rm{M}_\odot$, dust mass; \mdms, dust to stellar mass ratio; \ldust/$\rm{L}_\odot$, dust luminosity; \tauv, total effective $V$-band optical depth seen by stars in birth clouds; \tauvISM, the total effective $V$-band optical depth in the ambient ISM \sfr/$\rm{M}_\odot yr^{-1}$, the star-formation rate (SFR) averaged over the last $10^7$ years; and \ssfr/$yr^{-1}$, the specific star-formation rate (SSFR) averaged over the last $10^7$ years.
The ranges of each panel reflect the width of the priors for the SMG libraries. Where the prior range is different for the standard libraries the edge of the prior space is marked with a black dashed line.}
\label{fig:stacked_PDFs}
\end{figure*}

We show the median-likelihood physical parameters for each individual SMG with a good SED fit in Table~\ref{tab:all_SMG_properties}. To compare the physical parameters of the high and low redshift dusty populations, we compute the stacked probability density function (PDF) of parameters derived from the SED fitting, which are shown in Fig.~\ref{fig:stacked_PDFs}. For each parameter, we use the first moment of the stacked PDF to estimate the mean of the population, with the variance on the population taken from the second moment of the average PDF minus the mean squared. The error on the mean is simply the square root of the population variance, normalised by the square root of the number of galaxies in the sample. The mean values and errors on each PDF for the high and low redshift samples are summarised in Table \ref{tab:summary_properties}, including parameters for the SMGs derived using both set of priors. We show the mean PDF for the high-redshift SMG sample using the standard priors, to reassure the reader that the trends observed between the low and high-redshift samples are not driven by the use of different priors.

\subsection{Comparison of parameters for high and low redshift populations}
\label{Sample_params}
In this section we compare the mean physical parameters for the high redshift ($z>1$) SMGs and the low redshift mass-matched sample drawn from H-ATLAS. We note that using the whole low redshift H-ATLAS sample in place of the mass-matched sample produces a negligible difference in our results.

\begin{table*}
\begin{center}
\caption[Summary of the mean physical properties derived from stacking probability density functions.]{Summary of mean physical properties of the samples examined in this paper derived from stacking probability density functions (PDFs) for the different galaxy populations studied in this paper. The parameters are: \fmu, the fraction of total dust luminosity contributed by the diffuse ISM; ${M}_\ast/\rm{M}_\odot$, stellar mass; ${M}_\mathrm{d}/\rm{M}_\odot$, dust mass; \mdms, dust to stellar mass ratio; \ldust/$\rm{L}_\odot$, dust luminosity; \tauv, total effective $V$-band optical depth seen by stars in birth clouds; \tauvISM, effective $V$-band optical depth in the ambient ISM; $\sfr^{7}/\rm{M}_\odot yr^{-1}$, the star-formation rate (SFR) averaged over the last $10^7$ years; $\ssfr^{7}/yr^{-1}$, specific star-formation rate (SSFR) averaged over the last $10^7$ years; $\sfr^{8}/\rm{M}_\odot yr^{-1}$, the SFR averaged over the last $10^8$ years; and $\ssfr^{8}/yr^{-1}$, the SSFR averaged over the last $10^8$ years.}
\begin{tabular}{l D{,}{\pm}{8} D{,}{\pm}{8} D{,}{\pm}{8} }
\hline
  \multicolumn{1}{p{2.0cm}}{Parameter}
& \multicolumn{1}{p{2.5cm}}{\hfil Low redshift mass-matched sample}
& \multicolumn{1}{p{3.0cm}}{\hfil $z>1$ SMG sample}
& \multicolumn{1}{p{3.0cm}}{\hfil $z>1$ SMG sample} \\ 
  \multicolumn{1}{p{2.0cm}}{}
& \multicolumn{1}{p{3.0cm}}{\hfil (standard prior)}
& \multicolumn{1}{p{3.0cm}}{\hfil (SMG prior)}
& \multicolumn{1}{p{3.0cm}}{\hfil (standard prior)} \\ 
\hline      
$\fmu$ & 0.65,0.01 & 0.13,0.02 & 0.32,0.02 \\
log$_{10}$(${M}_\ast$)   & 10.73, 0.02 & 10.80, 0.10 & 10.97, 0.10 \\
log$_{10}$(${M}_\mathrm{d}$)   &  8.19, 0.02 &  9.09, 0.09 &  9.27, 0.07 \\
log$_{10}$($\mdms$)    & -2.54, 0.02 & -1.71, 0.10 & -1.70, 0.12 \\
log$_{10}$($\ldust$)    & 10.96, 0.02 & 12.57, 0.07 & 12.50, 0.07 \\
\tauv  &  2.7, 0.1 &  5.1, 0.6 &  3.4, 0.2 \\
\tauvISM  &  0.7, 0.1 &  1.0, 0.1 &  1.4, 0.1 \\
log$_{10}$($\sfr^7$)   & 0.51,0.03 & 2.59,0.08 & 2.50,0.08 \\
log$_{10}$($\ssfr^7$)     & -10.22,  0.03 &  -8.21,  0.11 &  -8.47,  0.12 \\
log$_{10}$($\sfr^8$)   & 0.56,0.03 & 2.53,0.08 & 2.23,0.07 \\
log$_{10}$($\ssfr^8$)  & -10.17,  0.03 &  -8.27,  0.10 &  -8.75,  0.09 \\
\hline                                                                                 
\end{tabular}                                                                                                      
\label{tab:summary_properties}
\end{center}
\end{table*}

\subsubsection*{Fraction of total dust luminosity contributed by the diffuse ISM: \fmu}
The dust luminosity in most SMGs is dominated by the birth cloud component, whilst the dust luminosity in low redshift galaxies is dominated by the diffuse ISM (Fig.~\ref{fig:stacked_PDFs}a). If the standard priors are used, the values of \fmu\, tend to be higher but we still find that the majority of the SMGs have \fmu$<0.5$. 

\subsubsection*{Stellar Mass: \mstar}
In Fig.~\ref{fig:stacked_PDFs}b we find a mean stellar mass of $6.3^{+1.6}_{-1.3}\times 10^{10}\,$M$_\odot$ for the $z>1$ SMGs, in agreement with \citet{Hainline11} and M12. Using \citet{BC03} models \citet{Michalowski12} found stellar masses for SMGs which were higher by a factor of $2-4$ compared to those in this study. This difference in stellar mass is due to the use of different SFHs, stellar population models and and the strength of the TP-AGB stars in the stellar population models. By design, the stellar mass of the low redshift mass-matched sample ($5.5\pm0.2 \times 10^{10}\,$M$_\odot$) is similar to that of the SMGs.

\subsubsection*{Dust Mass: \md}
The $z>1$ SMG sample has a mean dust mass of $1.2^{+0.3}_{-0.2}\times{10}^9\,$M$_\odot$ (Fig.~\ref{fig:stacked_PDFs}c), similar to other studies of SMGs \citep{Santini10, Magdis12, Simpson14}. The dust masses of the SMGs are around an order of magnitude higher than the low redshift H-ATLAS galaxies, which have a mean dust mass of $(1.6\pm0.1)\times{10}^8\,$M$_\odot$. Furthermore, there is a dearth of galaxies in the low redshift sample with dust masses as large as the dustiest SMGs (${M}_\mathrm{d}>2.5\times 10^{9}\,$M$_\odot$). It is not surprising that a high redshift submillimetre sample has a higher average dust mass, since moderate dust masses are not detectable at high redshifts with \emph{Herschel}. However, this selection effect does not account for the much larger space density of the dustiest galaxies at high redshift, since these would have been detected in H-ATLAS should they exist at lower redshift. This is consistent with the observed strong evolution in the dust content of massive, dusty galaxies with redshift, in agreement with \citet{DE01}, \citet{Dunne03b}, \citet{Eales10_Hermes}, \citet{Dunne11}, \citet{Bourne12a} and \citet{Symeonidis13}.

\subsubsection*{Dust-to-Stellar Mass: \mdms}
The \mdms values of $z>1$ SMGs in Fig.~\ref{fig:stacked_PDFs}(d) typically range from 0.01--0.05, with a mean of $0.019^{+0.005}_{-0.004}$, similar to that found by \citet{Santini10}. While \citet{Santini10} found that SMGs have a factor of 30 higher \mdms values compared to a sample of normal spirals from SINGS, we find our SMGs to be only a factor of 7 more dusty relative to their stellar mass compared to low redshift H-ATLAS galaxies. This disparity may be because \citet{Santini10} compare to a sample of very local galaxies, whereas the H-ATLAS sample is selected at $250\mu$m and covers a greater range in redshift, in which evolution in dust mass has already occurred \citep{Dunne11, Bourne12a}. 

\subsubsection*{Dust Luminosity: \ldust}
The dust luminosities of the low and high-redshift samples are significantly different (Fig.~\ref{fig:stacked_PDFs}e). The mean of the low redshift sample is $9.2^{+0.4}_{-0.3}\times 10^{10}$L$_\odot$, whereas the SMGs have an average dust luminosity a factor of 40 higher. The mean total dust luminosity of the high-redshift SMGs ($3.7^{+0.7}_{-0.6}\times 10^{12}$L$_\odot$) is in good agreement with M12.

\subsubsection*{Optical depth: \tauv, \tauvISM}
As shown in Fig.~\ref{fig:stacked_PDFs}(f) and (g), the total effective $V$-band optical depth seen by stars in birth clouds (\tauv) is is around a factor of two higher for the SMG sample compared to low redshift H-ATLAS galaxies, although the optical depth in the diffuse ISM (\tauvISM) is similar for the two samples. These results are consistent with other studies which found that SMGs are very obscured compared to local galaxies, but are not as obscured as local ULIRGs \citep{Menendez-Delmestre09}. This is the likely reason behind the higher \fmu\ values observed in SMGs. 

\subsubsection*{Star-formation rate: SFR}
The SFR of the SMGs (averaged over the last $10^7$ years) ranges from $62-2200\,$M$_\odot$\,yr$^{-1}$, but there is a strong trend of SFR with redshift. For SMGs at $z>1$ the mean is $390^{+80}_{-70}$  (Fig.~\ref{fig:stacked_PDFs}h) in agreement with other recent studies of similar samples \citep{Banerji12, LoFaro13, Simpson14}. We note that because we exclude 6 SMGs where the submillimetre emission may originate from multiple sources at the same redshift, the sample may be biased against systems undergoing major mergers, which tend to have the highest SFRs. The average SFR of the SMGs is around 120 times that of the low redshift sample (SFR $=3.3\pm{0.2}$\,M$_\odot$yr$^{-1}$). The lack of highly star-forming galaxies in the low redshift sample is not a volume effect, as the co-moving volume probed by the H-ATLAS Phase 1 data is $1.1\times10^{8}$\,Mpc$^{3}$, which is comparable to the co-moving volume of the SMG sample from the combined SPIRE survey areas of GOODS-N, ECDFS, COSMOS and Lockman Hole ($1.3\times10^{8}$\,Mpc$^{3}$ for $1.0<z<5.31$). Submillimetre selected galaxies at fixed stellar mass have higher SFRs at higher redshift, which reflects the strong evolution in characteristic SFR in galaxies out to $z\sim2$ \citep{Sobral13}.

\subsubsection*{Specific Star-Formation rate: SSFR}
The mean SSFR of the $z>1$ SMG sample in Fig.~\ref{fig:stacked_PDFs}i is $6.1^{+1.7}_{-1.3}\times 10^{-9}$\,yr$^{-1}$, which implies a doubling time of $160$\,Myr. The SSFR values derived from our SED fitting are in broad agreement with those from M12 derived from the FIR luminosity, albeit with large scatter. The average SSFR of the SMGs from the {\sc magphys} SED fitting is around 100 times greater than the mean SSFR of the low redshift sample, which has an average SSFR of $6.1^{+0.5}_{-0.4}\times 10^{-11}$\,yr$^{-1}$. The difference in the mean PDFs when using the SMG and standard priors for the SFR and SSFR averaged over the last $10^7$ years are 0.09 and 0.26\,dex, respectively. When using the SMG priors these results are not sensitive to the timescale over which the (S)SFR is averaged, although with the standard priors the mean SMG (S)SFR is lower when averaging over a longer timescale of $10^8$ years. This is due to the birthcloud timescale being fixed at $10^7$ years in the standard model, which is unable to generate the high optical depths (and hence obscured star-formation rates) required to fit all of the SMG SEDs. However, the choice of prior or timescale over which to average SFR does not change the conclusion that dusty galaxies at high redshift are forming more stars than dusty galaxies of a similar stellar mass at low redshift.

\subsection{SEDs of dusty galaxies at low and high redshift}
\label{Sec:stacked_SEDs}

\begin{figure}
\centering
\includegraphics[width=0.48\textwidth]{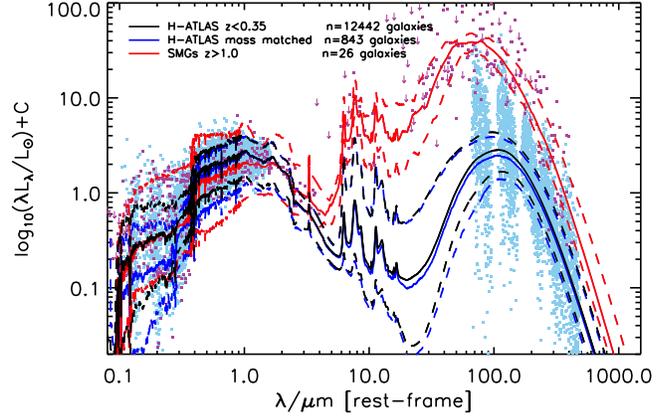}
\caption{Median stacked SEDs of the entire H-ATLAS sample with $z<0.35$ based on the updated \citet{Smith12} H-ATLAS SED fits (black), the low redshift mass-matched sample (blue) and the $z>1$ SMGs (red). The thick lines show the median of the best-fit {\sc magphys} SEDs, and the dotted lines show the $1\sigma$ spread around the median SED. The red and blue points show the deredshifted and normalised observed photometry for the SMGs and the low redshift mass-matched sample. The SEDs are normalised at rest-frame $2.2\mu$m. The stellar mass-matched sample is broadly consistent with the stack of the full H-ATLAS sample at wavelengths $>4000$\AA. The stacked SMG SED is much more obscured, hot and luminous compared to the low redshift H-ATLAS SED.} 
\label{fig:stacked_SEDs_all}
\end{figure}

We now investigate the shapes of the SEDs of the galaxies in our sample. In Fig.~\ref{fig:stacked_SEDs_all} we show the median SEDs of all H-ATLAS galaxies, mass-matched H-ATLAS galaxies and $z>1$ SMGs. The median SEDs are derived using a similar method to that presented \citet{Smith12}, but with $\sim10$ times as many sources. Since we are comparing stacked SEDs of a similar stellar mass, the SEDs are normalised at $2.2\mu$m. In Fig.~\ref{fig:stacked_SEDs_all} the median SED of the stellar mass-matched sample is broadly consistent with the stack of the full H-ATLAS sample at wavelengths $>4000$\AA. At shorter wavelengths, the mass-matched sample is redder which is most likely due to it sampling the highest mass end of the H-ATLAS distribution which has a greater contribution from lower SSFR objects (see Fig~\ref{fig:stacked_PDFs}). The MIR region in the H-ATLAS stacked SED of shows the largest variation, as each best-fit SED is only weakly constrained by the model priors, however, \citet{Smith12} showed that a lack of MIR data does not affect the results derived from the SED fitting. The stacked SMG SED is much more obscured, hot and luminous compared to the low redshift H-ATLAS SED. Whilst we note that we could be biased towards warmer SMGs in our sample due to the need for at least one PACS/SPIRE detection, \citet{Magnelli12} have shown that the dominant selection bias in the SMG sample is due to the need for (sub)mm and radio detections. This striking visual confirmation of the shift in SED shape was implied by \citet{Lapi11} to occur in the submillimetre selected population in the interval $0.5 < z < 1.5$.

\begin{figure*}
\begin{minipage}[t]{1.0\textwidth}
\begin{center}
$\begin{array}{cc}
\includegraphics[width=0.48\textwidth]{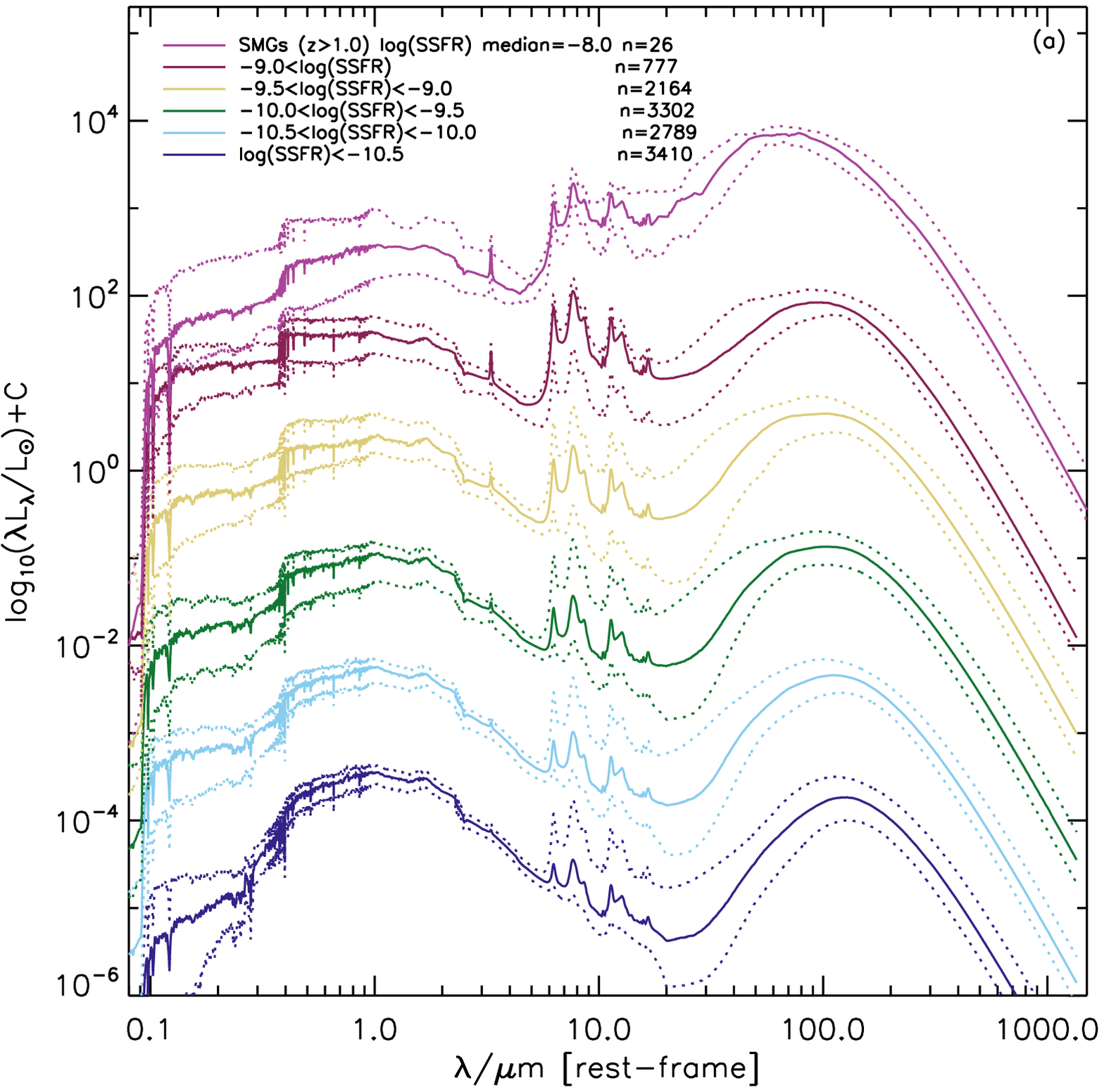} &
\includegraphics[width=0.48\textwidth]{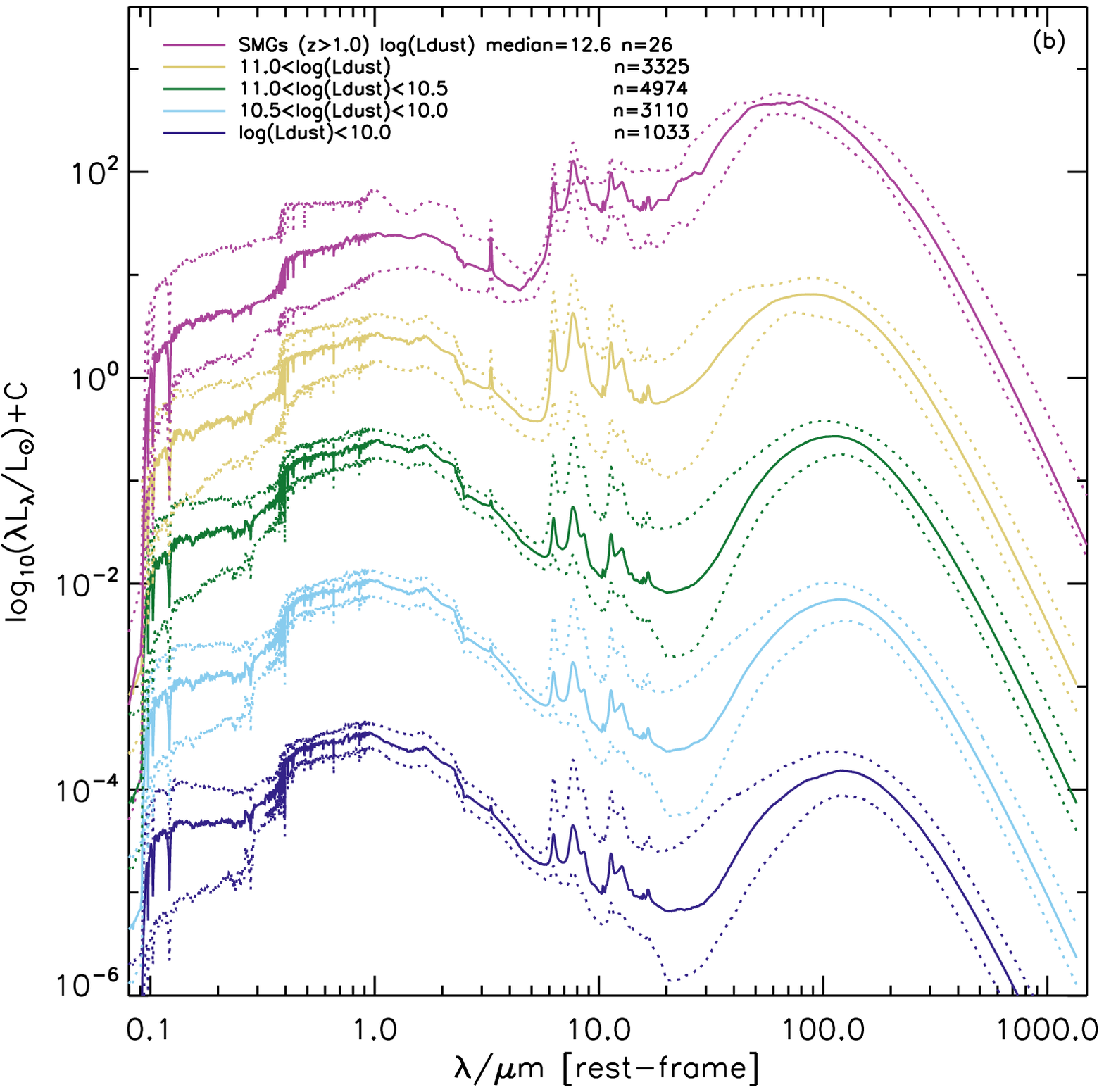} \\ 
\end{array}$
\end{center}
\end{minipage}
\caption{(a): Median stacked SEDs in bins of best-fit SSFR for the $z>1$ SMGs (upper red SED) and the entire $z<0.35$ H-ATLAS sample, based on the updated \citet{Smith12} H-ATLAS SED fits. The SEDs become bluer in the optical and have a hotter dust continuum with increasing SSFR. The stacked SMG SED shows a redder optical continuum due to increased obscuration compared to the H-ATLAS galaxies. (b): Stacked SEDs in bins of best-fit dust luminosity. The SEDs have a redder optical continuum with increasing \ldust, and increase in dust temperature for L$>10^{11}\rm{L_{\odot}}$. The dotted lines show the $1\sigma$ spread around the median SED.}
\label{fig:stacked_SEDs}
\end{figure*}

\begin{figure*}
\begin{minipage}[t]{1.0\textwidth}\
\begin{center}
$\begin{array}{cc}
\includegraphics[width=0.40\textwidth]{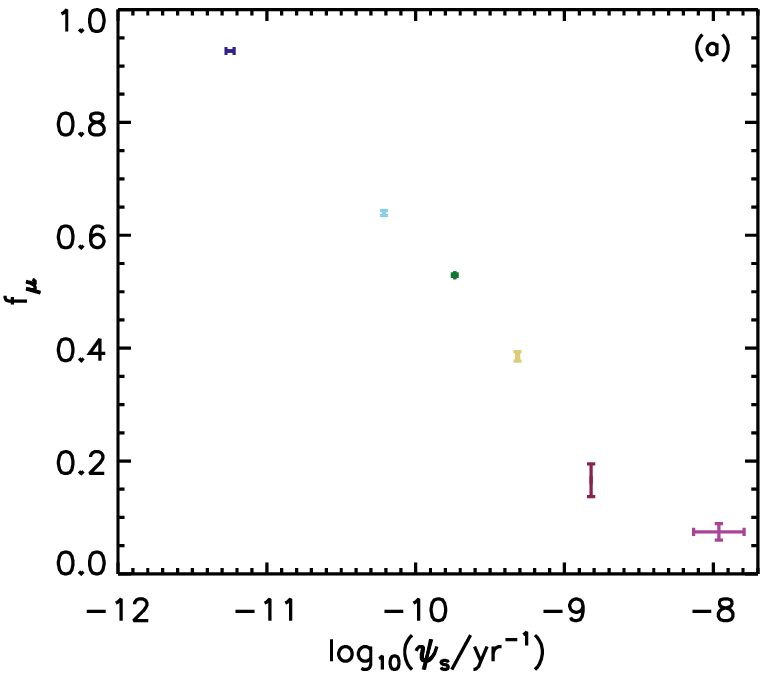} &
\includegraphics[width=0.40\textwidth]{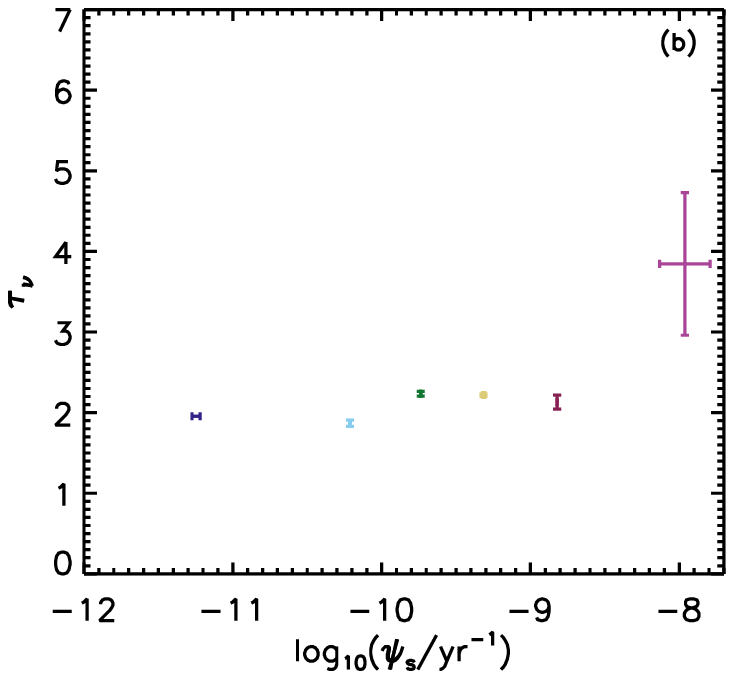} \\ 
\end{array}$
\end{center}
\end{minipage}
\caption{The relationship between the median values of the best-fit \fmu\, and SSFR (a) and \tauv\, and SSFR (b) in each SSFR bin in Fig.~\ref{fig:stacked_SEDs}. Error bars indicate the standard error on the median for each bin using the method in \citet{Gott01}. The average \fmu\, for the SMGs (red point) does not show a sharp change compared to the \fmu\, values of the low redshift sample in bins of SSFR. The values of \tauv\, are relatively constant with increasing SSFR for the low redshift H-ATLAS galaxies, but there is a sharp increase in \tauv\, for the SMGs.}
\label{fig:fmu_ssfr}
\end{figure*}

In Fig.~\ref{fig:stacked_SEDs} we show the median SEDs of the $z>1$ SMGs and the updated H-ATLAS empirical SED templates from \citet{Smith12}, binned by best-fit SSFR and dust luminosity. 
In Fig.~\ref{fig:stacked_SEDs}(a), there is a strong trend for the SEDs of $z<0.35$ H-ATLAS galaxies to become bluer in the optical with increasing SSFR and hotter in the dust continuum (see also \citealt{Smith12}). However, the SMG bin (which has minimal overlap in SSFR with the low-redshift H-ATLAS galaxies) shows quite a break in the optical--UV trend, with the SMG SED being much redder and more obscured. The trend for warmer dust continuum continues, together with a marked increase in the ratio of IR to optical-UV continuum. Thus for a modest increase in SSFR, the stacked SMG SED looks very different to the most actively star-forming galaxies at $z<0.35$ in H-ATLAS. Most H-ATLAS galaxies have \fmu\, values which indicate around half of their \ldust\, is contributed by birth clouds, while SMGs have much lower values of \fmu\, suggesting $\geq 80$ percent of their \ldust\, is produced in obscured star-forming regions. 
The change in SED shape could be due to SMGs having more birth cloud relative to diffuse ISM luminosity. In Fig.~\ref{fig:fmu_ssfr}(a) we see a steady decrease in the value of \fmu\, as SSFR increases, such that the highest SSFR bin for $z<0.35$ H-ATLAS sources has a similar \fmu\, to the SMGs. The sudden change in the optical--UV SED between the highest SSFR H-ATLAS galaxies and the SMGs cannot be due to a sharp change in \fmu; rather it must be due to a physical difference in the structure of birth clouds in SMGs. Our SED fitting prefers that the birth clouds in SMGs have a higher optical depth on average (see Fig.~\ref{fig:fmu_ssfr}b), and the stars are also able to spend longer in them (suggesting they last longer before disruption). We return to this subject in \S\ref{sec:SFE}.

For SEDs binned by dust luminosity in Fig.~\ref{fig:stacked_SEDs}(b), the SEDs show a more steady trend of becoming redder in the optical with increasing dust luminosity and also warmer in the infrared above a dust luminosity of $\sim 10^{11}\rm{L_{\odot}}$ (consistent with \citealt{Smith12}). There is no marked difference in the trend once the SMG bin is reached, as is apparent for the SSFR binning.

\subsection{\label{sec:Lfir} Infrared luminosity as a star-formation tracer in submillimetre-selected galaxies}

\begin{figure}
\centering
\includegraphics[width=0.48\textwidth]{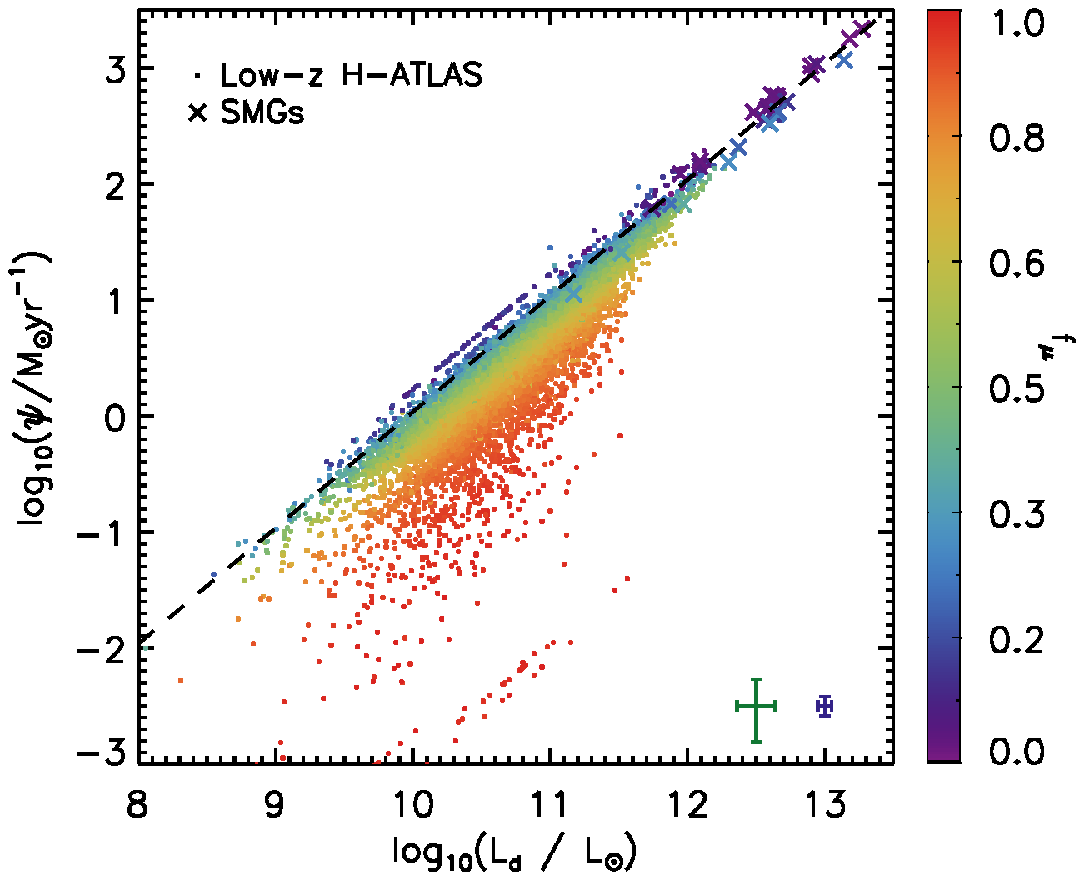}
\caption{The relation between median likelihood dust luminosity and star-formation rate
  for the low redshift H-ATLAS galaxies (dots) and SMGs
  (crossed). Points are coloured by the value of \fmu. The dashed
  line shows the relation between SFR and total infrared luminosity (integrated from $8-1000\mu$m)
  from \citet{Kennicutt98}. The error bars indicate the median 84th--16th percentile range from each individual parameter PDF; the green and blue error bars correspond to the low redshift H-ATLAS and SMG samples, respectively. Galaxies with low \fmu\ lie on the K98 relation, which means that for SMGs this relation is a reliable way
of predicting the SFR from the total infrared luminosity. For galaxies with a significant contribution to the infrared luminosity from the diffuse ISM the K98 relation will produce an over-estimate of the SFR from \ldust.}
\label{fig:LIR_SFR}
\end{figure}

Studies of infrared and submillimetre selected galaxies have traditionally
relied on using the re-radiated energy from dust at $8-1000\mu$m as a proxy
for SFR. The seminal work by \citet[][hereafter K98]{Kennicutt98} explains in detail the basis of this relationship and provides a calibration (see also \citealt{Kennicutt09}). The main requirement for dust luminosity to be a good tracer of SFR is that the bulk of the star formation is obscured and the dust emission is produced from absorption of photons produced by massive stars\footnote{There is a slightly different calibration \citep{Hao11} if UV emission is being added to the infrared luminosity in order to capture both the obscured and unobscured component.}. Since MAGPHYS aims to account for both obscured (radiated in the FIR) and unobscured (radiated in the UV) star
formation, {\em and\/} also accounts for that fraction of \ldust\ which
is heated by older stellar populations, it is instructive to look at
the correlation of the MAGPHYS SFR with \ldust~(Fig.~\ref{fig:LIR_SFR}). Galaxies with low \fmu\ lie on the K98 relation, which means that for SMGs, this relation is a reliable way
of predicting the SFR from the total infrared luminosity -- as expected
given their high obscuration. Galaxies with a significant contribution
to the infrared luminosity from the diffuse ISM (mostly powered by
stars older than 10\,Myr) lie further from the K98 relation, and are mostly low-redshift H-ATLAS galaxies. 
Using \ldust\ and the K98 relation will therefore \emph{over-estimate the SFR in
galaxies} where the dust luminosity is produced mainly in the diffuse
ISM component (i.e. high \fmu). The robustness of MAGPHYS SFR relative
to a number of well used SFR tracers is investigated further in a
study by Smith et al. (in prep).

\subsection{\label{sec:SFE} Understanding the ISM in SMGs and low redshift galaxies}
The mass of dust and SFR are correlated in
galaxies \citep{dC10}; such a relationship might be expected if dust is a tracer of
the gas content in galaxies \citep{Eales12, Scoville13}, as gas mass and SFR are
linked by the Kennicutt-Schmidt relation \citep{Kennicutt98}. To investigate this
idea we show in Fig.~\ref{fig:high-redshift_mstar_properties}~(a) the SFR
from MAGPHYS versus the dust mass. In these plots we also include local ULIRGS fitted using MAGPHYS in the study by \citet{dC10b}. The $z>1$ SMGs and local ULIRGs follow a parallel but offset relationship
from the H-ATLAS $z<0.5$ sources. Interestingly, the $z<1$ SMGs lie closer to the H-ATLAS sources than to the other SMGs. If dust is a good tracer of gas, this implies that high-redshift SMGs and local ULIRGS have \emph{more SFR per
unit gas mass} than the $z<0.5$ H-ATLAS galaxies. The quantity $\rm{SFR}/\md$ is therefore inversely proportional to a gas depletion timescale, $\tau_g$, (or proportional to a star-formation efficiency), under the assumption of a roughly uniform gas-to-dust ratio for galaxies in this sample. Fig.~\ref{fig:high-redshift_mstar_properties}(a) implies shorter gas depletion timescales for high-redshift SMGs and ULIRGs than for more `normal' galaxies at $z<0.5$ \citep{Tacconi08, Genzel10}.

Such differences between SMGs and `normal' star forming galaxies have been found in previous studies of gas and SFR which use CO to trace the molecular gas \citep{Tacconi08, Dannerbauer09, Genzel10, Daddi10b}, but see also \citet{Ivison11}. Fitting to the samples in Fig~\ref{fig:high-redshift_mstar_properties}~(a) for the
$z<0.5$ H-ATLAS galaxies, and the $z>1$ SMGs\footnote{Where we keep the slope fitted to the SMGs/ULIRG sample the same as the low-redshift sample.} and local ULIRGs gives:

\begin{equation}
\begin{tabular}{>{\raggedright\arraybackslash}m{4.6cm} >{\raggedright\arraybackslash}m{3.5cm}}
\label{eqn:sfr_mdust}
$\log_{10} \rm{SFR} = 1.16 \log_{10} \md - 7.81$ & for $z>1$ SMGs\\
 & \quad \& low-z ULIRGs.\\ 
$\log_{10} \rm{SFR} = 1.16 \log_{10} \md - 8.72$ & for $z<0.5$ H-ATLAS.\\ 
\end{tabular}
\end{equation}

\noindent These have the same slope as that fitted to the \lfir{\footnote{Integrated from $50-300\mu$m.} versus
\lco\ relationship of \citet{Genzel10} (hereafter G10). Since \lfir\,
should be proportional to SFR for the sources in G10 (see
\S\ref{sec:Lfir}), we can infer that dust mass appears to
trace molecular gas (for galaxies selected in the submillimetre) at least as
well as \lco. To convert \lco\
into a mass of molecular hydrogen we must assume a conversion factor
($\alpha_{\rm{CO}}$), which depends on the dynamical state of the gas,
and potentially also the metallicity \citep[G10,][]{Leroy11, Narayanan12, Sandstrom13}. Galaxies
with strong nuclear starbursts, or which are mergers (e.g. local ULIRGs) are found to
often have a lower $\alpha_{\rm{CO}}$ \citep{Solomon97, Downes_Solomon98, Yao03} due to their gas being in a smoother, more diffuse state; no longer acting like an
ensemble of virialised self-gravitating clouds\footnote{An assumption which underlies the 'standard' conversion from \lco to $M_{\rm {H_2}}$.}. Typically authors have used the lower `local ULIRG' value, $\alpha_{CO} = 0.8 - 1.0 \,\rm{M_{\odot}\, (K\,km\,s^{-1}\,pc^{2})^{-1}}$ when studying high-redshift SMGs, under the assumption that their high infrared luminosities are also powered by compact starbursts, leading to similar conditions in the gas. While this appears to be appropriate in many cases \citep{Tacconi06, Tacconi08, Magdis11, Magnelli12b}, there are significant caveats about using it `wholesale' for any SMG \citep{Ivison11, Papadopoulos12b, Bothwell13}. In particular, the latter authors warn that the mass of dense gas in these systems may be underestimated when using the standard ULIRG value for $\alpha_{\rm{CO}}$ and when only using lower excitation CO lines ($J<3$). 

In light of these outstanding issues, we will treat the conversion of CO luminosity to gas mass as an uncertain step and highlight any impacts of choosing a particular value of $\alpha_{\rm{CO}}$ on our conclusions. Using an $\alpha_{\rm{CO}} = 0.8 - 3.2 \,\rm{M_{\odot}\, (K\,km\,s^{-1}\,pc^{2})^{-1}}$
for SMGs/ULIRGS and normal star forming galaxies respectively, we
  can translate \lco\ in the G10 relationship to
  $M_{\rm{H_2}}$, using $M_{\rm{H_2}} = 1.36 \, \alpha_{\rm{CO}}$ \lco \,
  \msun (where the factor 1.36 accounts for the mass of Helium). We
  translate the y-axis of the G10 relation using the K98
  relationship: $L_{\rm{IR}} = 10^{10} \, \rm{SFR}$ for a Chabrier IMF, and
  follow G10 in converting \lfir\, to
  \lir \footnote{Integrated from $8-1000\mu$m.} with a factor of 1.3. We can thus express the G10
    relationships as:

\begin{equation}
\begin{tabular}{>{\raggedright\arraybackslash}m{4.9cm} >{\raggedright\arraybackslash}m{3.5cm}}
\label{eqn:cosfr}
$\log_{10} \rm{SFR} = 1.15 \log_{10} M_{\rm{H_2}} - 9.30$ & for $z>1$ SMGs\\
 & \quad \& low-z ULIRGs\\ 
$\log_{10} \rm{SFR} = 1.15 \log_{10} M_{\rm{H_2}} - 10.60$ & for SFGs.\\ 
\end{tabular}
\end{equation}

\noindent We now have a relationship between SFR and dust mass in our samples
and a relationship between SFR and molecular gas (as traced by CO)
from G10 for comparable samples of SMGs/ULIRGs and
normal star-forming galaxies. At a given gas-to-dust ($G_d$) ratio,
these two relationships (SFR vs $M_{\rm H_2}$ and SFR vs \md) will be
  equivalent. This happens at $G_d = 30 - 150$ for SMGs (depending on
  the choice of $\alpha_{\rm{CO}}$) and $G_d = 80$ for normal star forming
  galaxies (the $z<0.5$ H-ATLAS sample). These values are
  consistent with observations of high-redshift SMGs \citep{Kovacs06, Swinbank14} and star-forming galaxies in the local Universe \citep{Seaquist04, Draine07, Leroy11, Cortese12, Sandstrom13}.
  
Not only does this comparison suggest that dust is as good a tracer of
molecular gas as CO, but the consistency of the implied gas-to-dust
ratios with observations of gas and dust in individual objects also suggests that the dust masses from MAGPHYS are reasonable and that evolution in $\kappa_d$, the dust mass
absorption coefficient, is not responsible for the shift in the SMGs
relative to the H-ATLAS sources in Fig~\ref{fig:high-redshift_mstar_properties}~(a). In fact, for a change in $\kappa_d$ to explain this shift, the dust masses of the SMGs and ULIRGs would
need to be higher by a factor 5 to bring them onto the same relation
as the $z<1$ galaxies. This would produce extremely high \mdms
values and very low inferred $G_d$ from observations of CO, none of which are physically sensible given chemical and dust evolution modelling (Rowlands et al. 2014, submitted).

In Fig.~\ref{fig:high-redshift_mstar_properties}~(b) we plot \mdms\ as a
function of SSFR, which essentially normalises the first plot by stellar
mass so that we can compare the `specific' quantities. The
addition of SMGs allows us to extend the investigation of the
\mdms--SSFR relation to higher redshifts, beyond that studied in
\citet{dC10} and \citet{Smith12}. Again using \md\ as a proxy for gas mass, the $y$-axis (\mdms) is proportional to $f_g/(1-f_g)$, where $f_g$ is the baryonic gas fraction ($f_g=M_g/[M_g+M_{\ast}]$). Galaxies at the same horizontal position in
this figure are thus equally `gas-rich'\footnote{Under the assumption they have the same average gas-to-dust ratio.}. From Fig.~\ref{fig:high-redshift_mstar_properties}~(b) it is clear that SMGs are on average more `gas rich' than the lower-redshift H-ATLAS galaxies {\em and\/} the local ULIRGs (this agrees with detailed studies using CO, e.g. \citealp{Tacconi06, Geach11, Bothwell13}). It is also apparent that the local ULIRGs have significant overlap in gas fraction with the H-ATLAS galaxies.

There is once again a significant offset in the locus of the $z>1$ SMGs and local ULIRGs compared to the lower redshift H-ATLAS galaxies, such that {\em at the same gas fraction, SMGs/ULIRGs have more star-formation activity than `normal' star forming submillimetre selected sources\/}. This is an important point as it means that `gas richness' alone cannot explain the offset between the samples in Fig.~\ref{fig:high-redshift_mstar_properties}~(a) -- something else must happen in SMGs to push them into a more rapid and efficient conversion of their gas supply into stars. Recalling the change in optical SEDs between normal star forming galaxies and SMGs from \S\ref{Sec:stacked_SEDs}, it is likely that the physical changes in the ISM which lead to enhanced star-formation efficiency are also the cause of the increased obscuration in the UV/optical. 
Observations of local ULIRGs have shown that high density gas components ($N_{\rm H_2}>10^5\,\rm{cm^{-3}}$) are dominant \citep{Gao_Solomon04}, and this is thought to be responsible for their high star-formation efficiencies \citep{Greve09}. At high redshift, galaxies are generally more gas rich (as we see from Fig~\ref{fig:high-redshift_mstar_properties}b), and simulations of turbulent gas-rich disks have shown that they are dynamically unstable to fragmentation and collapse on a large scale \citep{Elmegreen_Burkert10}. This situation occurs on smaller scales in local ULIRGs, but which requires a major merger to initiate the instability in local galaxies \citep{BarnesHernquist91, Mihos_Hernquist94, Mihos_Hernquist96, Barnes_Hernquist96}. 

The star forming clumps in high-redshift SMGs are distributed over larger spatial scales ($\sim2$\,kpc) than those in local ULIRGs ($50-200$\,pc), though the physical conditions inside them appear to be similar \citep{Swinbank11}. The clumps appear in simulations and can last for $\sim 10^8$\,yr, possibly due to the higher pressure in the ISM in SMGs/ULIRGs and high redshift gas-rich systems \citep{Genzel08, Swinbank11, Bournaud14}. Such large, dense and long lived star forming regions may be the reason for the high obscuration in these systems (recall that we needed to adjust the birth-cloud timescale parameter in MAGPHYS to achieve good fits). While mergers at high redshift will certainly produce the instability required to promote the collapse of the disk into large and dense clumps \citep{Bournaud11, Hayward11}, it is not clear that they are necessary in all cases. 
    
\begin{figure}
\centering
\includegraphics[width=0.48\textwidth, height=18cm, trim=3mm 1mm 1mm 1mm]{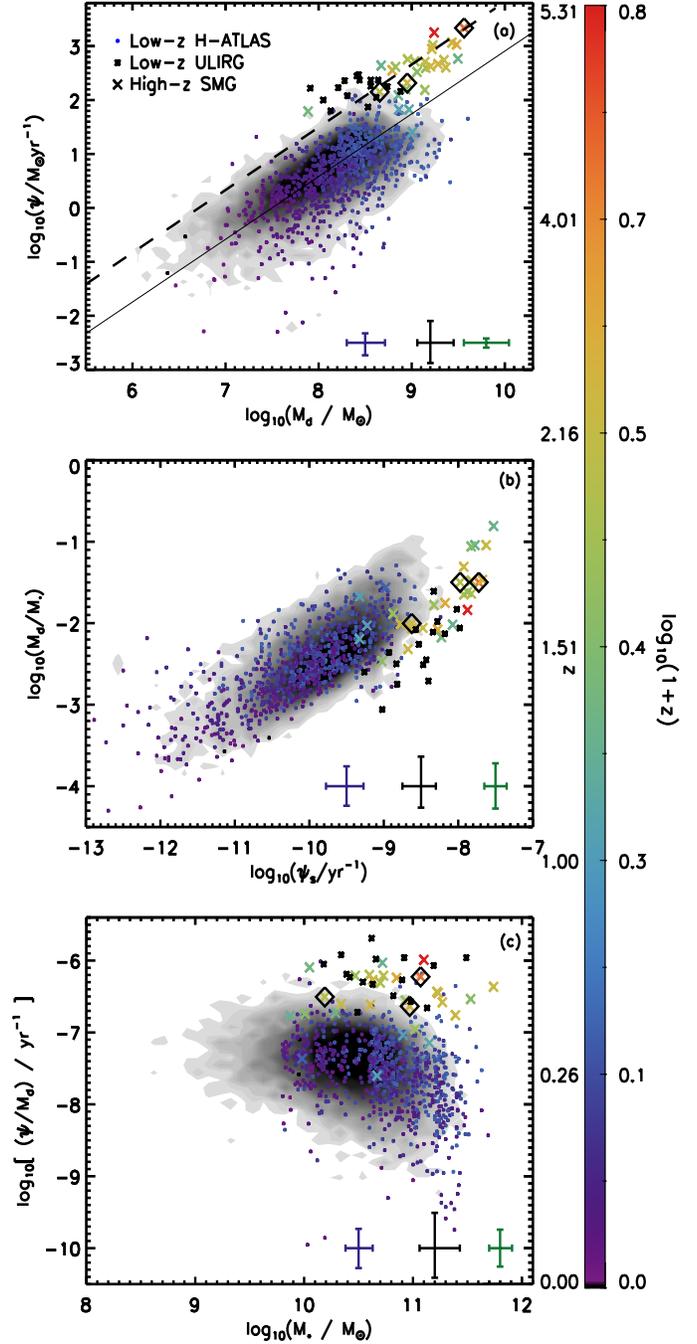}
\caption{(a) The relation between median likelihood SFR and dust mass,
  (b) dust-to-stellar mass ratio and SSFR, and (c) SFR/\md\, and
  stellar mass for SMGs, the mass-matched low-redshift sample and
  low redshift ULIRGs (crosses, dots and black stars
  respectively). The shaded contours show the locus of the main
  H-ATLAS sample. The SFR and SSFR are averaged over the last $10^7$
  years. Points are coloured by redshift. Open black diamonds indicate the three SMGs in our sample whose FIR photometry may be confused. The error bars indicate the median 84th--16th percentile range from each individual parameter PDF; the blue, black and green error bars correspond
  to the low redshift H-ATLAS, ULIRG and SMG samples,
  respectively. In panel (a) the solid line is the fit to the H-ATLAS sample, and the dashed line is the fit to the $z>1$ SMGs and low redshift ULIRGs, keeping the slope fixed to that of the H-ATLAS sample.}
\label{fig:high-redshift_mstar_properties}
\end{figure}

Recalling that $\rm{SFR}/\md$ is inversely proportional to a gas depletion timescale, Figure~\ref{fig:high-redshift_mstar_properties}~(c) shows our proxy for 1/$\tau_g$ (or star-formation efficiency) as a function of stellar
mass. The star-formation efficiency of SMGs and local ULIRGs show no trend with stellar mass, and
have much shorter gas-depletion timescales (higher star-formation efficiencies) than the low-redshift sample at all stellar masses. Using the mean SFR and dust mass of SMGs from \S~\ref{Sample_params} and the $G_d$ inferred from equating our relationships with those of G10 we estimate $\tau_g \sim 90-470$\,Myr for the $z>1$ SMGs. The gas-depletion timescales of the $z<1$ SMGs are consistent with the low-redshift
H-ATLAS galaxies, which have $\tau_g \sim 4$\,Gyr (using $G_d=80$ inferred from Eqn.~ \ref{eqn:sfr_mdust} and \ref{eqn:cosfr}). The low-redshift H-ATLAS sample shows a slight trend, such that more massive galaxies have longer gas-depletion timescales. Some
of these low efficiency galaxies are passive and not actively forming
stars \citep{Rowlands12}, however, removing all sources with SSFR $<10^{-11}$ yr$^{-1}$
does not change the overall trend. The relation between star-formation efficiency and stellar mass mirrors that seen between SSFR and stellar mass for the $z<0.5$ galaxies.

To explain the offset between the SMGs and H-ATLAS galaxies as a result of metallicity differences (and therefore gas-to-dust ratio changes) would
require evolution of the mass-metallicity relationship of order a dex
or more from $z=0.5$ to $z\sim 2-3$. This evolution in metallicity is not observed \citep{Mannucci10, Stott13}. Since similar offsets between star-forming galaxies and SMGs are found in studies which rely on CO as a gas mass tracer \citep{Tacconi06, Tacconi08, Genzel10, Daddi10b, Bothwell13}, we conclude that these differences between these galaxy populations are genuine.

\subsection{The nature of star formation in SMGs}
MAGPHYS also produces a best-fit SFH for each galaxy, which is normalised to reproduce the best-fit stellar mass from the SED fit. While these SFHs are not unique solutions (see \citet{Rowlands14a} for a discussion), it is still instructive to see which mode of star formation is fitted in these sources. Fig.~\ref{fig:SFHs} shows the SFHs of the SMG sample. Most of them could be described as `bursts' of star formation, either because they have a short elevated SFR near the current age, or because their SFHs are so short and extreme they can be considered a burst. The same conclusion was found by \citet{dC10b} in their study of local ULIRGs. Notably the $z<1$ SMGs are those with the least current star formation in the SMG sample and had their last burst some time ago, consistent with their similarity to the $<0.5$ H-ATLAS galaxies. As expected, SMGs are therefore likely to rapidly exhaust their gas supply within a few hundred Myr \citep[][and references within]{Simpson14}.

\section{Conclusions}
\label{sec:conclusions}
We have presented the physical properties and SEDs of a rest-frame $250\mu$m selected sample of massive, dusty galaxies, in the range $0<z<5.3$. The sample consists of a compilation of 29 high-redshift SMGs with photometry from \citet{Magnelli12} and 843 dusty galaxies at $z<0.5$ from the \emph{Herschel}-ATLAS, selected to have a similar stellar mass to the SMGs. Both samples have panchromatic photometry from the rest-frame UV to the submillimetre, which allowed us to fit SEDs to derive statistical constraints on galaxy physical parameters using an energy balance technique. We compared the physical properties of the high and low redshift samples and found significant differences in the submillimetre-selected galaxy populations. Our main results are as follows:

\begin{itemize}

\item The sample of $z>1$ SMGs have an average SFR of $390^{+80}_{-70}\,$M$_\odot$yr$^{-1}$ which is around 120 times that of the low redshift sample matched in stellar mass to the SMGs (SFR $=3.3\pm{0.2}$\,M$_\odot$yr$^{-1}$). This is consistent with the observed evolution in characteristic SFR of galaxies out to $z\sim2$. The SMGs harbour an order of magnitude more dust ($1.2^{+0.3}_{-0.2}\times{10}^9\,$M$_\odot$), compared to $(1.6\pm0.1)\times{10}^8\,$M$_\odot$ for low redshift dusty galaxies selected to have a similar stellar mass.

\item From the SED analysis we find that a large fraction of the dust luminosity in SMGs originates from star-forming regions, whereas at lower redshifts the dust luminosity is dominated by the diffuse ISM. The means that for SMGs the SFR can be reliably predicted from the K98 calibration between far-infrared luminosity and SFR. Where the dust luminosity is produced mainly by the diffuse ISM component, the \citet{Kennicutt98} relation will over-estimate the SFR, which is the case for the majority of low redshift H-ATLAS galaxies.
 
\item The median SED of the SMGs is more luminous, has a higher effective temperature and is more obscured, with stars in birth clouds experiencing a factor of $\sim2$ more obscuration compared to the median low redshift H-ATLAS SED. There is a sudden change in the optical--UV SED between the highest SSFR H-ATLAS galaxies and the SMGs, which cannot be due to a sharp change in the contribution to the total dust luminosity from birth clouds. Since the effective optical depth in SMGs is higher than in H-ATLAS galaxies the change in SED shape may be due to a physical difference in the structure of birth clouds in SMGs.

\item We find that at the same dust mass the SMGs are offset by 0.9\,dex towards a higher SFR compared to the low redshift H-ATLAS galaxies. This is not only due to the higher gas fraction in SMGs but also because they are undergoing a more efficient mode of star formation. The offset cannot be explained by differences in the metallicities between the samples, or variations in the dust emissivity.

\item The offset in SFR and dust mass between the SMGs and low redshift galaxies is similar to that found in CO studies. Due to the consistency between observations of gas and dust in individual SMGs and the gas-to-dust ratios implied by the ratio of FIR to CO luminosity we conclude that dust mass is as good a tracer of molecular gas as CO.

\item At the same gas fraction, SMGs/ULIRGs have more star-formation activity than `normal' star-forming $250\mu$m selected sources. This is consistent with their best-fit SFHs which are bursty in nature.

\end{itemize}

\section*{Acknowledgements}
We thank the referee for useful suggestions which have much improved the clarity of this manuscript.
We thank Michal Micha{\l}owski, Malcolm Bremer, Shane Bussmann and Asantha Cooray for helpful comments.
K.~R. acknowledges support from the European Research Council Starting
Grant SEDmorph (P.I. V.~Wild).
I.R.S. acknowledges support from the SFTC (ST/I001573/1), a Leverhulme Trust fellowship, the ERC Advanced Grant Dustygal and a Royal Society/Wolfson Merit Award. 
We acknowledge the use of print-friendly colour tables by Paul Tol (http://www.sron.nl/~pault/).
The Herschel-ATLAS is a project with \emph{Herschel}, which is an ESA space
observatory with science instruments provided by European-led
Principal Investigator consortia and with important participation from
NASA. The H-ATLAS website is http://www.h-atlas.org/. GAMA is a
joint European-Australasian project based around a spectroscopic
campaign using the Anglo-Australian Telescope. The GAMA input
catalogue is based on data taken from the Sloan Digital Sky Survey and
the UKIRT Infrared Deep Sky Survey. Complementary imaging of the GAMA
regions is being obtained by a number of independent survey programs
including \emph{GALEX} MIS, VST KIDS, VISTA VIKING, \emph{WISE}, \emph{Herschel}-ATLAS,
GMRT and ASKAP providing UV to radio coverage. GAMA is funded by the
STFC (UK), the ARC (Australia), the AAO, and the participating
institutions. The GAMA website is http://www.gama-survey.org/.
This research has made use of data from HerMES project (http://hermes.sussex.ac.uk/). 
HerMES is a \emph{Herschel} Key Programme utilising Guaranteed Time from the SPIRE instrument team, ESAC scientists and a mission scientist. HerMES is described in \citet{Oliver10_Hermes}.
This research has made use of the NASA/ IPAC Infrared Science Archive, which is operated by the Jet Propulsion Laboratory, California Institute of Technology, under contract with the National Aeronautics and Space Administration.

\bibliographystyle{mn2e}
\bibliography{references}

\appendix

\section{Standard and SMG priors}
\label{sec:Comparison of priors}

Here we highlight the parameters which are different in the standard and SMG prior libraries. A summary of standard and SMG prior distributions are shown in Fig.~\ref{fig:priors}.

\begin{figure*}
\centering
\includegraphics[width=1.0\textwidth, height=22cm]{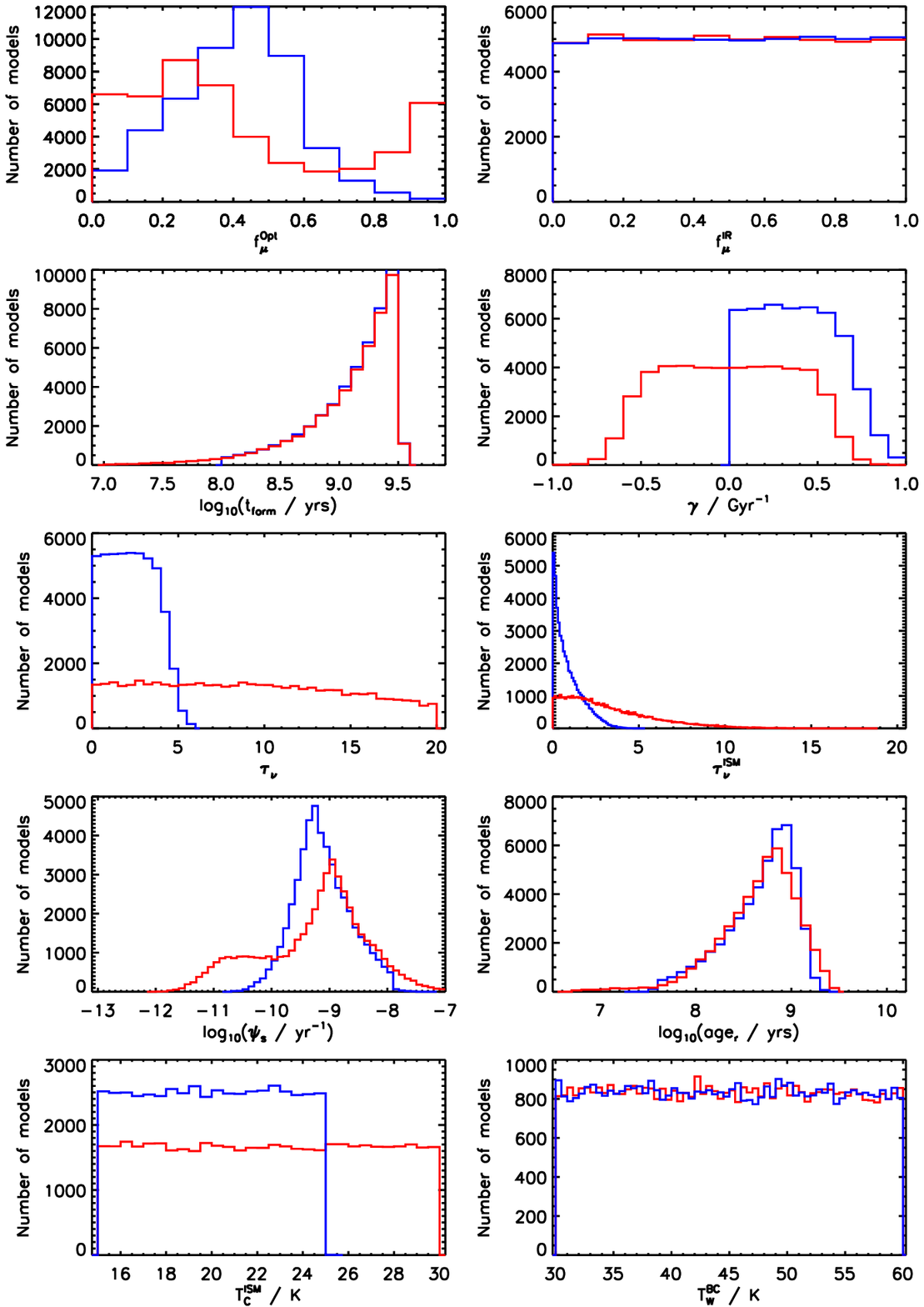}
\caption[Comparison of priors.]{Comparison of the standard (blue histogram) and SMG (red histogram) prior distributions for parameters relevant to this work at $z\sim2$. The panels are: \fmuOpt, the fraction of total dust luminosity contributed by the diffuse ISM in the optical model; \fmuIR, the fraction of total dust luminosity contributed by the diffuse ISM in the infrared model; $\gamma$, the star-formation timescale (Gyr$^{-1}$); \tauv, total effective $V$-band optical depth seen by stars in birth clouds; \tauvISM, the total effective $V$-band optical depth in the ambient ISM; \ssfr/$yr^{-1}$, specific star-formation rate, $\mathrm{age_{r}}$, $r$-band light-weighted age; \tbgscold/K, temperature of the cold diffuse ISM dust component; and \tbgswarm/K, temperature of the warm dust component in birth clouds.}
\label{fig:priors}
\end{figure*}

\paragraph*{Optical Depth:}
From DCE08, the standard priors for \tauv\, and \tauvISM, the total effective $V$-band optical depth seen by stars in birth clouds and in the ambient ISM, respectively, range from 0 to 6. This describes the full range of attenuations observed for normal low redshift galaxies (DCE08, and references within). When fitting the SEDs of SMGs with the standard priors, the \tauv\, PDF frequently runs up against the upper edge of the prior space. This suggests that the \tauv\, prior in the standard libraries does not extend to sufficiently high values to fully describe the properties of SMGs (which are known to be more obscured than local galaxies \citealp{Menendez-Delmestre09}). As with the ULIRG priors in \citet{dC10b}, the \tauv\, and \tauvISM priors are modified to allow for higher optical depths so that they now range between 0 and 20.

\paragraph*{Star-formation history (SFH):}
The standard prior for the SFH is parametrised as an exponentially decreasing model of the form exp$(-\gamma t)$, where $\gamma$ is the star-formation time-scale parameter and is distributed uniformly between 0 and 1 Gyr$^{-1}$. For SMGs, we also adopt both exponentially increasing and decreasing star-formation rates by distributing the $\gamma$ parameter as a  Gaussian between -1 and 1 Gyr$^{-1}$, as many studies \citep{Maraston10, Lee10, Papovich11, Reddy12b} found that an exponentially increasing SFR may be appropriate for some high-redshift galaxies.  

Bursts are superimposed at random times on the continuous SFH, but with a probability that 50 per cent of galaxies experience a burst in the last 2 Gyr. The strength of the burst is defined as the mass of stars formed in the burst relative to the mass of stars formed in continuous star formation over the lifetime of the galaxy. This parameter ranges from 0.03 -- 4.0 with logarithmic spacing in the standard prior. Since SMGs are thought to be experiencing strong starbursts, the burst strength is increased to range from 0.1 -- 100. 

The time since the start of star formation in the galaxy ($t_\mathrm{form}$) is uniformly distributed between 0.1 and 13.5 Gyr in the standard prior. The lower limit is decreased from 0.1 to 0.01 Gyr in the SMG prior in order to increase both the number of models with SSFR $\sim 1\times{10}^{-8}$~yr$^{-1}$ and to extend the upper limit of the SSFR prior from $\sim1\times{10}^{-8}$~yr$^{-1}$ to $\sim1\times{10}^{-7}$~yr$^{-1}$.

\paragraph*{Birth Cloud Timescale:}
Moderately star-forming galaxies in the local Universe are assumed in the DCE08 model to have a fixed birth cloud timescale ($t_\mathrm{BC}$) of $1\times10^7$ years, after which the young stars move from their birth clouds into the less obscured diffuse ISM. \citet{dC10b} found that $t_\mathrm{BC}$ = $1\times10^8$ years was more appropriate for ULIRGs, which are more heavily obscured than normal star-forming galaxies. For the SMGs, we allowed $t_\mathrm{BC}$ to vary as a free parameter which is uniformly distributed in logarithmic space between $1\times10^7$ and $1\times10^8$ years. This accounts for the possibility of longer birth cloud lifetimes in gas-rich disks \citep{Krumholz10} but does not force SMGs to have such extreme opacities as local ULIRGs.

\paragraph*{Dust temperatures:}
The temperature of the cold dust component is extended from $15-25$\,K to $15-30$\,K, as was done for the ULIRGs in \citet{dC10b}. The greater intensity of star formation in the SMGs could produce higher ambient dust temperatures in the ISM, due to an increase in the hardness of the interstellar radiation field.

\subsection{Comparison of priors}
\label{sec:Comparison_of_priors}
Figure~\ref{fig:compare_composite2_standard_libs} shows the parameter values derived using the standard {\sc magphys} and SMG libraries for the 29 high-redshift SMGs with good SED fits (of which four are AGN power-law subtracted SEDs). The parameters which appear to be most sensitive to the choice of prior are \fmu\,, \tauv\ and SSFR, which is not unexpected given this was the aim of altering the priors. Increasing \tauv\ and the birthcloud timescale in the SMG priors results in more of the dust luminosity (which is constrained very well by observations) being produced in the birth clouds in the model. Around 50 per cent of our sources have SSFRs significantly higher than would be obtained using the standard priors. 
Parameters where we did not specifically alter the prior (e.g. \md\,, SFR, \mstar) are reassuringly not very different. There is a slight tendency for stellar masses and dust masses to be lower with the SMG priors, this is a systematic change but still within the median error range for these parameters. In the majority of cases the choice of prior for the SMGs does not change our conclusions in \S\ref{sec:results}. Where the choice of prior influences our results we highlight the effect when interpreting our findings.

\begin{figure*}
\centering
\includegraphics[width=1.0\textwidth]{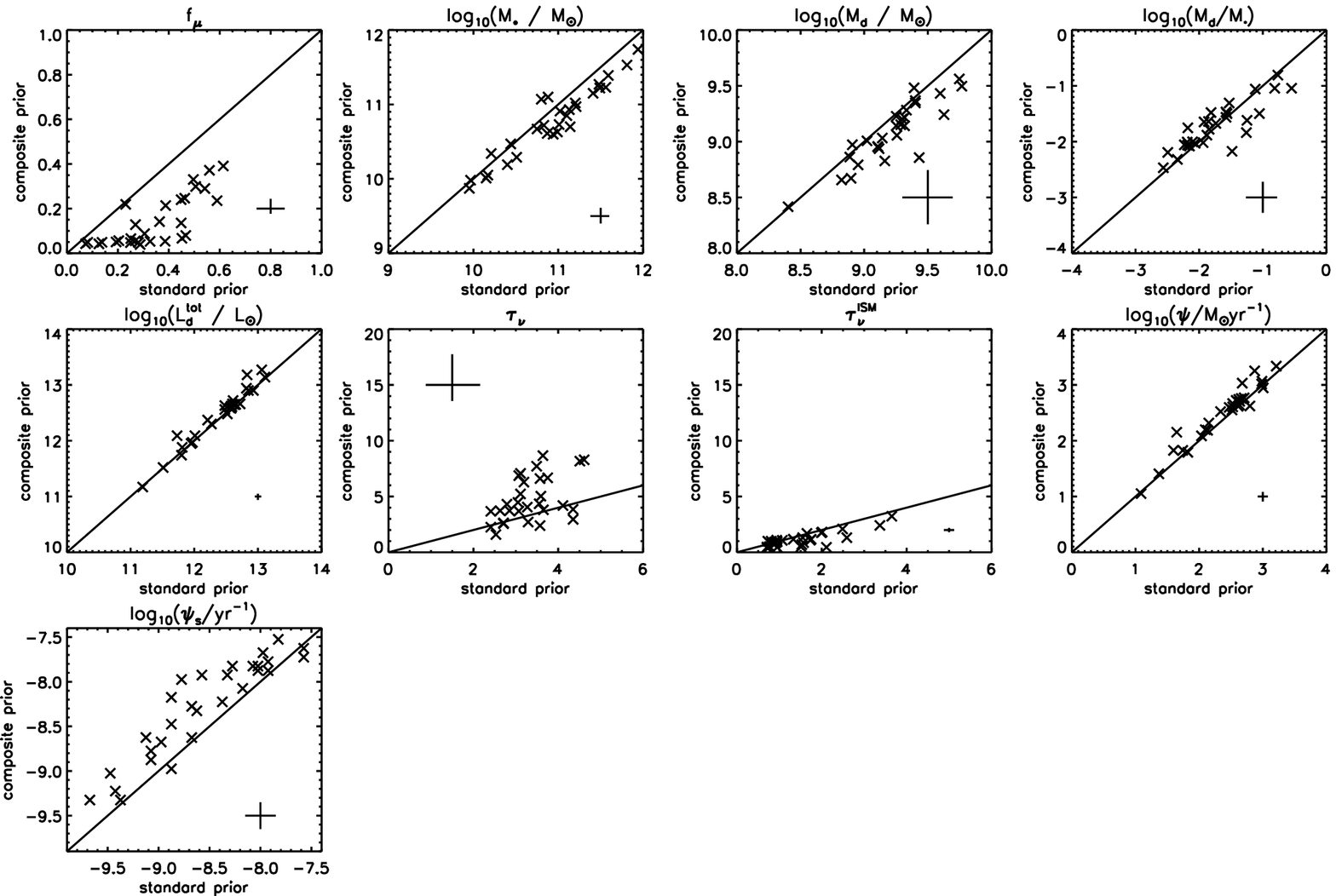}
\caption[A comparison of the median-likelihood values of different parameters using the standard {\sc magphys} prior libraries and the new SMG priors.]{A comparison of median-likelihood values of different parameters using the standard {\sc magphys} prior libraries and the new SMG priors for the 29 submillimetre galaxies in \S\ref{sec:results}. Note that where the axis ranges are different these reflect the width of the priors. Solid black lines show the one-to-one line for each parameter. The error bar indicates the median 84th--16th percentile range from the parameter PDF. The parameters shown are: \fmu, the fraction of total dust luminosity contributed by the diffuse ISM; ${M}_\ast/{M}_\odot$, stellar mass; ${M}_\mathrm{d}/{M}_\odot$, dust mass; \mdms, dust-to-stellar mass ratio; \ldust/${L}_\odot$, dust luminosity; \tauv, total effective $V$-band optical depth seen by stars in birth clouds; \tauvISM, the total effective $V$-band optical depth in the ambient ISM; \sfr/${M}_\odot yr^{-1}$, the star-formation rate (SFR); and \ssfr/$yr^{-1}$, the specific star-formation rate averaged over the last $10^7$ years.}
\label{fig:compare_composite2_standard_libs}
\end{figure*}

\section{SED fits}
We present the panchromatic SED fits for the sample of SMGs studied in this paper, using the SMG {\sc magphys} priors described in \S\ref{sec:SED_fitting}.

\begin{figure*}
\centering
\includegraphics[width=1.0\textwidth]{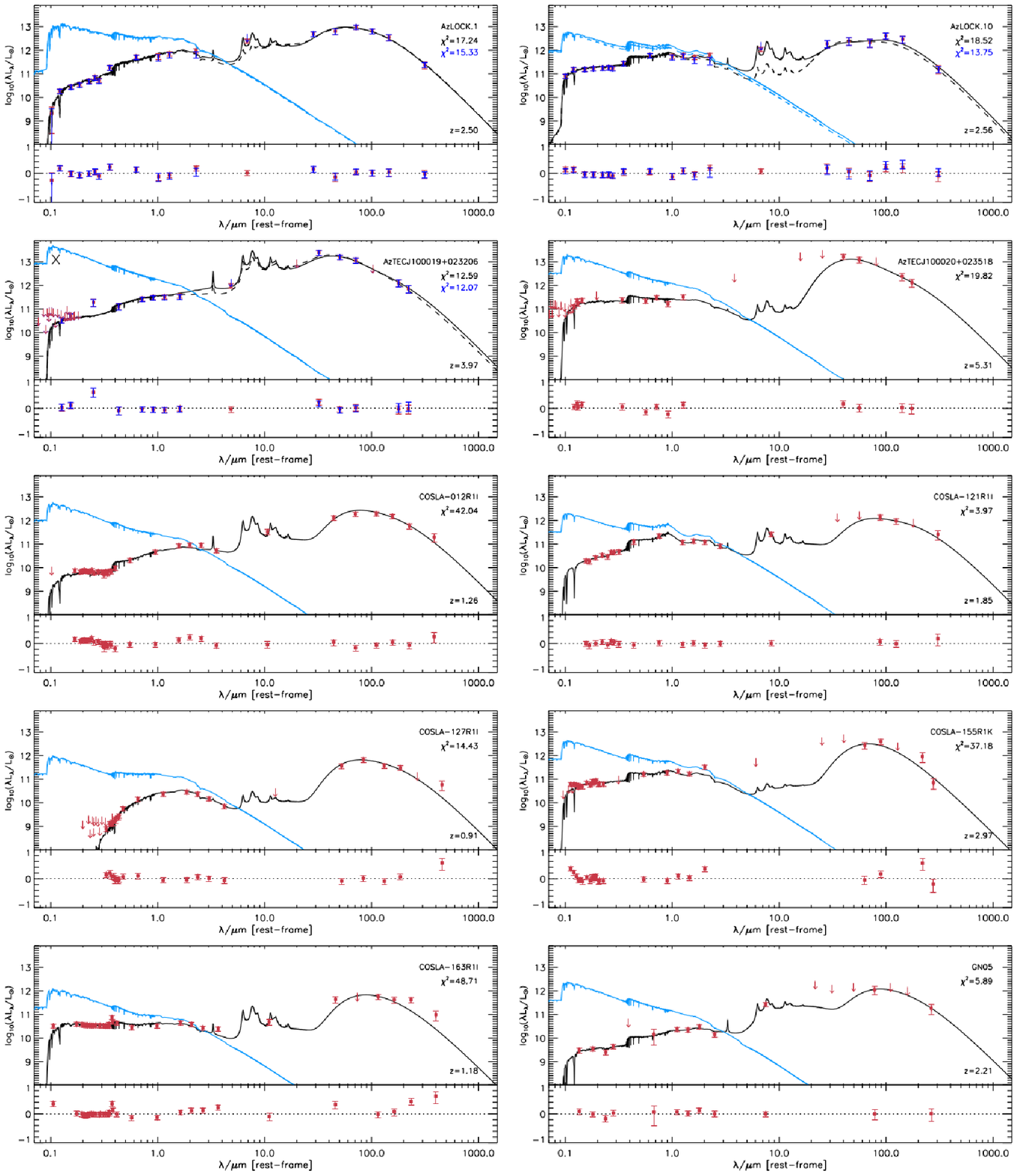}
\caption{Multiwavelength SEDs of the 34 SMGs in our sample (including five rejected fits indicated by a black cross in the top left corner of each plot), with observed photometry (red points) from the rest-frame UV to the submillimetre. Upper limits are shown as arrows, and errors on the photometry are described in \S\ref{sec:High_redshift_sample}. The solid black line is the best-fit model SED and the solid blue line is the unattenuated optical model. The residuals of the fit are shown in the panel below each SED. In the case where we have subtracted a power-law component to account for hot dust emission from an AGN, the dashed lines indicate the best-fit model, and the blue points indicate the power-law subtracted photometry.}
\label{fig:SEDs}
\end{figure*}

\begin{figure*}
\centering
\includegraphics[width=1.0\textwidth]{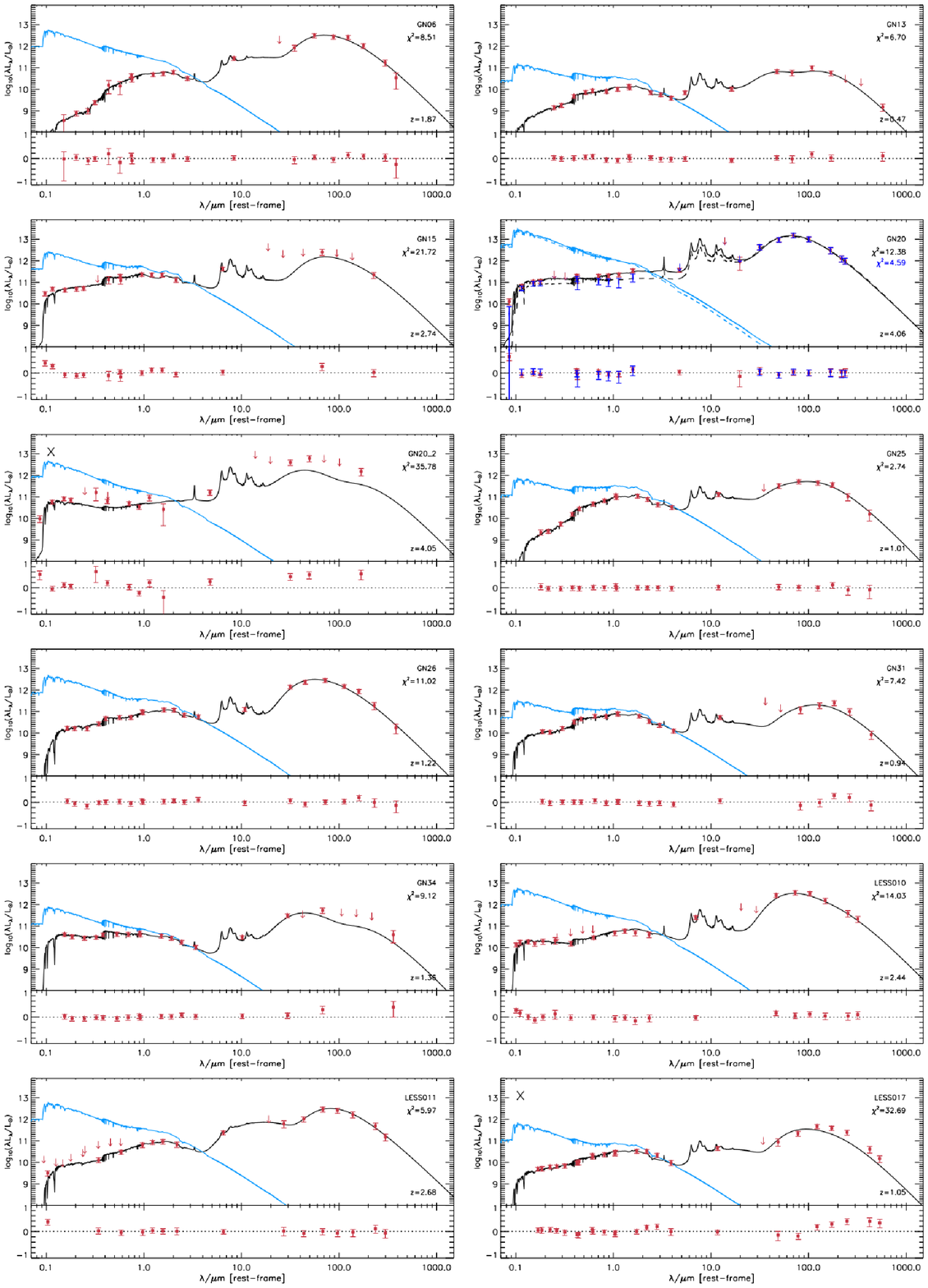}
\contcaption{}
\end{figure*}

\begin{figure*}
\centering
\includegraphics[width=1.0\textwidth]{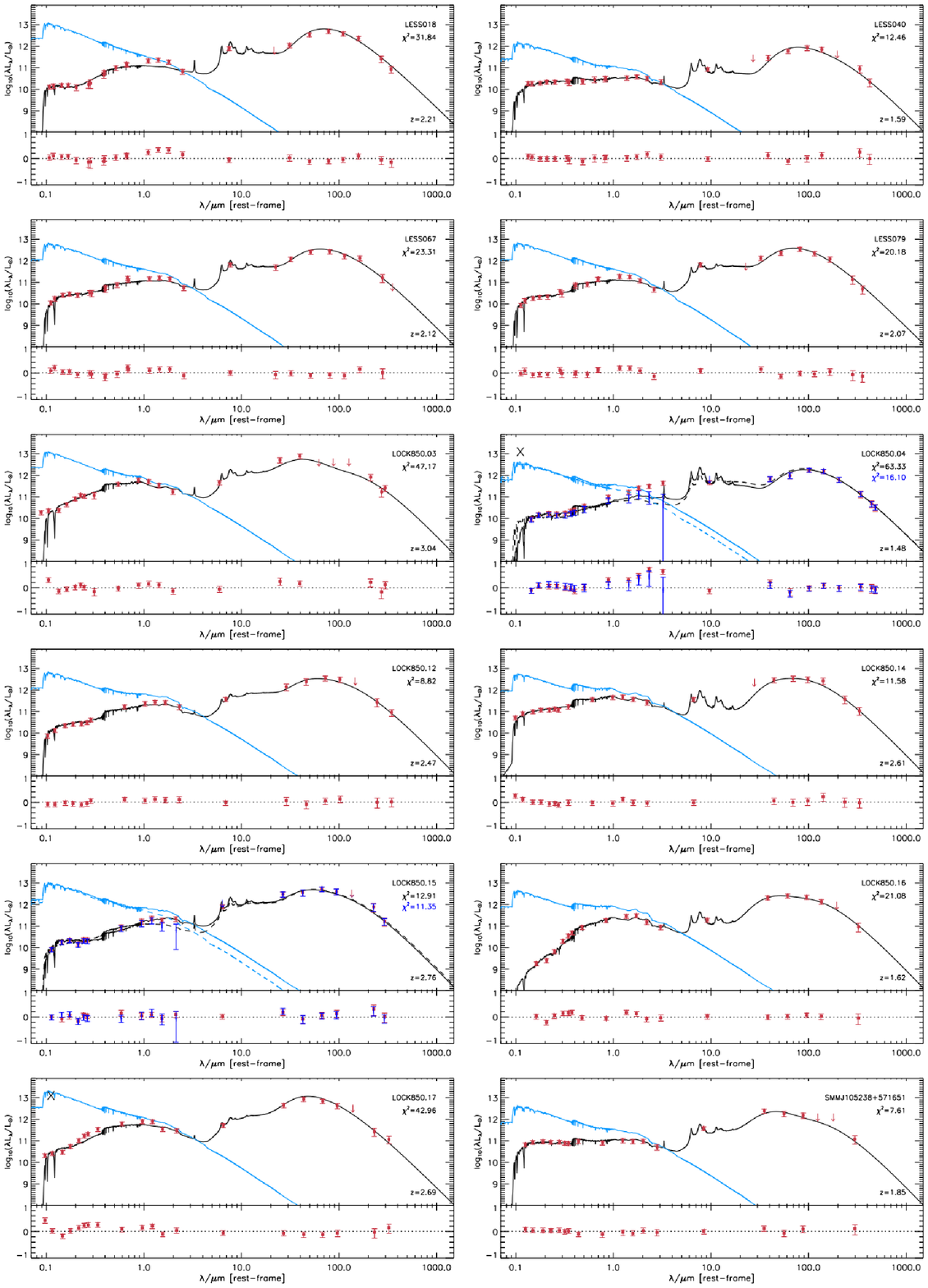}
\contcaption{}
\end{figure*}

\section{AGN}
\label{sec:AGN}
Some SMGs in our sample show excess emission in the rest-frame NIR, which may be due to dust heated to high temperatures by an obscured AGN \citep{Hainline11}. The {\sc magphys} SED models do not include a prescription for AGN emission and so we must assess the impact that AGN emission may have on the parameters. 

To select galaxies from our SMG sample which have power law emission in the NIR, we use the $S_{24}/S_{8.0}-S_{8.0}/S_{4.5}$ diagram from \citet{Ivison04} and the colour cut $S_{8.0}/S_{4.5}>1.65$ from \citet{Coppin10}. We find 6/34 galaxies at z>1 are classified as AGN (AzLOCK.01, AzLOCK.10, AzTECJ100019+023206, GN20, LOCK850.04 and LOCK850.15). We note that our sample may be slightly biased towards AGN because of the requirement of strong emission lines in order to measure a spectroscopic redshift.

We quantify the effect of power law emission on the derived physical parameters of SMGs selected to have a NIR excess. Following the method in \citet{Hainline11}, we parametrise the NIR excess emission as a simple power law with $f_{\lambda} \propto \lambda^{\alpha}$. The power law parametrisation does not include any prescription for dust extinction. We use values of $\alpha=2$ and 3 which are appropriate for SMGs \citep[][and references within]{Hainline11}. We normalise the power law to the observed $8\mu$m data point which is the maximum power law fraction. We then subtract from all photometry shortwards of $8\mu$m the power law flux in increments of $0.1\times$ the maximum power law fraction. We fit the power law subtracted SED at each increment to determine the galaxy physical parameters. The power law contribution to each galaxy SED is determined as the combination of power law and stellar emission model from {\sc magphys} which results in the best-fitting SED. Examples of the power law subtraction method are shown in Fig.~\ref{fig:SEDs}. When the power law fraction is large the optical emission can be over-subtracted; in this case we set the flux density to an upper limit at the value of the power law. There is evidence to suggest that the power law slope in the MIR is different to that in the optical--NIR, and \citet{Hainline11} found that extrapolation of the NIR power law longwards of $8\mu$m does not give a good prediction of the $24\mu$m flux density. Given the uncertainty in the AGN contribution to the MIR emission, we include data with $5<\lambda_\mathrm{rest}<30\mu$m as an upper limit in the SED fitting procedure. We assume that photometry longwards of rest-frame $30\mu$m has a negligible contribution from AGN emission \citep{Netzer07, Hatziminaoglou10, Pozzi12}.

\section{Physical properties of SMGs}

In Table~\ref{tab:all_SMG_properties} we show the median-likelihood physical parameters for each individual SMG derived from the {\sc magphys} SED fitting, as described in \S\ref{sec:SED_fitting}.

\begin{table*}
\begin{center}
\caption{Properties of the 29 SMGs (those with good SED fits) derived from SED fitting. The columns are (from left to right): Name, redshift, \fmu, the fraction of total dust luminosity contributed by the diffuse ISM; ${M}_\ast/\rm{M}_\odot$, stellar mass; ${M}_\mathrm{d}/\rm{M}_\odot$, dust mass; \mdms, dust to stellar mass ratio; \ldust/$\rm{L}_\odot$, dust luminosity; \tauv, total effective $V$-band optical depth seen by stars in birth clouds; \tauvISM, the total effective $V$-band optical depth in the ambient ISM \sfr/$\rm{M}_\odot yr^{-1}$, the star-formation rate (SFR) averaged over the last $10^7$ years and \ssfr/$yr^{-1}$, the specific star-formation rate (SSFR) averaged over the last $10^7$ years. Uncertainties are indicated by the median 84th--16th percentile range from each individual parameter PDF. Parameters derived from an AGN power-law subtracted SED are denoted with a \dag.}
\begin{tabular}{ >{\raggedright\arraybackslash}m{2.9cm} >{\centering\arraybackslash}m{0.5cm} >{\centering\arraybackslash}m{1.25cm} >{\centering\arraybackslash}m{1.25cm}> {\centering\arraybackslash}m{1.0cm} >{\centering\arraybackslash}m{1.2cm} >{\centering\arraybackslash}m{1.2cm}>{\centering\arraybackslash}m{1.1cm} >{\centering\arraybackslash}m{1.1cm}>{\centering\arraybackslash}m{1.0cm} >{\centering\arraybackslash}m{1.0cm} }
\hline
Name & $z$ & \fmu & \mstar & \md & \mdms & \ldust &\tauv & \tauvISM & \sfr & \ssfr \\ 
\hline   
\vspace{0.2cm}
AzLOCK.1\dag & 2.50 & $ 0.24^{+ 0.07}_{- 0.09}$ & $11.74^{+ 0.13}_{- 0.10}$ & $ 9.43^{+ 0.18}_{- 0.17}$ & $-2.32^{+ 0.22}_{- 0.20}$ & $13.14^{+ 0.05}_{- 0.05}$ & $ 6.67^{+ 5.38}_{- 2.80}$ & $ 1.75^{+ 0.22}_{- 0.26}$ & $3.07^{+0.08}_{-0.08}$ & $ -8.68^{+  0.15}_{-  0.20}$ \\
AzLOCK.10\dag & 2.56 & $ 0.25^{+ 0.06}_{- 0.07}$ & $11.39^{+ 0.06}_{- 0.08}$ & $ 9.37^{+ 0.25}_{- 0.25}$ & $-2.01^{+ 0.26}_{- 0.28}$ & $12.65^{+ 0.07}_{- 0.07}$ & $ 3.73^{+ 3.52}_{- 1.44}$ & $ 0.85^{+ 0.08}_{- 0.16}$ & $2.61^{+0.08}_{-0.08}$ & $ -8.77^{+  0.15}_{-  0.15}$ \\
AzTECJ100020+023518 & 5.31 & $ 0.04^{+ 0.03}_{- 0.02}$ & $11.10^{+ 0.06}_{- 0.10}$ & $ 9.24^{+ 0.52}_{- 0.30}$ & $-1.84^{+ 0.52}_{- 0.33}$ & $13.18^{+ 0.11}_{- 0.10}$ & $ 7.07^{+ 0.76}_{- 3.12}$ & $ 0.83^{+ 0.12}_{- 0.18}$ & $3.25^{+0.11}_{-0.08}$ & $ -7.88^{+  0.10}_{-  0.01}$ \\
COSLA-012R1I & 1.26 & $ 0.06^{+ 0.02}_{- 0.02}$ & $10.29^{+ 0.01}_{- 0.01}$ & $ 9.50^{+ 0.28}_{- 0.24}$ & $-0.81^{+ 0.26}_{- 0.24}$ & $12.61^{+ 0.01}_{- 0.01}$ & $ 3.85^{+ 0.01}_{- 0.01}$ & $ 0.45^{+ 0.01}_{- 0.01}$ & $2.77^{+0.01}_{-0.01}$ & $ -7.53^{+  0.01}_{-  0.01}$ \\
COSLA-121R1I & 1.85 & $ 0.30^{+ 0.07}_{- 0.09}$ & $11.02^{+ 0.16}_{- 0.10}$ & $ 9.14^{+ 0.48}_{- 0.38}$ & $-1.88^{+ 0.49}_{- 0.40}$ & $12.30^{+ 0.05}_{- 0.09}$ & $ 3.73^{+ 4.76}_{- 1.38}$ & $ 1.17^{+ 0.14}_{- 0.14}$ & $2.19^{+0.09}_{-0.10}$ & $ -8.88^{+  0.20}_{-  0.15}$ \\
COSLA-127R1I & 0.91 & $ 0.24^{+ 0.21}_{- 0.11}$ & $10.91^{+ 0.27}_{- 0.24}$ & $ 8.86^{+ 0.38}_{- 0.28}$ & $-2.03^{+ 0.38}_{- 0.40}$ & $11.88^{+ 0.06}_{- 0.07}$ & $ 8.17^{+ 4.58}_{- 0.92}$ & $ 2.41^{+ 1.02}_{- 0.20}$ & $1.83^{+0.12}_{-0.20}$ & $ -9.23^{+  0.45}_{-  0.15}$ \\
COSLA-155R1K & 2.97 & $ 0.13^{+ 0.08}_{- 0.07}$ & $10.85^{+ 0.11}_{- 0.04}$ & $ 8.79^{+ 0.22}_{- 0.22}$ & $-2.08^{+ 0.24}_{- 0.24}$ & $12.56^{+ 0.15}_{- 0.23}$ & $ 4.47^{+ 2.74}_{- 2.10}$ & $ 0.93^{+ 0.12}_{- 0.24}$ & $2.55^{+0.17}_{-0.25}$ & $ -8.27^{+  0.15}_{-  0.35}$ \\
COSLA-163R1I & 1.18 & $ 0.06^{+ 0.02}_{- 0.03}$ & $ 9.87^{+ 0.16}_{- 0.01}$ & $ 8.86^{+ 0.26}_{- 0.24}$ & $-1.04^{+ 0.26}_{- 0.24}$ & $11.95^{+ 0.01}_{- 0.01}$ & $ 2.25^{+ 2.18}_{- 0.01}$ & $ 0.41^{+ 0.16}_{- 0.01}$ & $2.09^{+0.01}_{-0.09}$ & $ -7.78^{+  0.01}_{-  0.15}$ \\
GN05 & 2.21 & $ 0.07^{+ 0.07}_{- 0.05}$ & $10.19^{+ 0.11}_{- 0.24}$ & $ 8.66^{+ 0.41}_{- 0.49}$ & $-1.50^{+ 0.40}_{- 0.51}$ & $12.09^{+ 0.18}_{- 0.17}$ & $ 5.03^{+ 3.38}_{- 1.88}$ & $ 1.17^{+ 0.46}_{- 0.42}$ & $2.15^{+0.17}_{-0.21}$ & $ -7.97^{+  0.25}_{-  0.30}$ \\
GN06 & 1.87 & $ 0.14^{+ 0.07}_{- 0.10}$ & $10.93^{+ 0.12}_{- 0.14}$ & $ 9.15^{+ 0.16}_{- 0.15}$ & $-1.78^{+ 0.20}_{- 0.20}$ & $12.62^{+ 0.03}_{- 0.04}$ & $ 8.29^{+ 5.36}_{- 3.46}$ & $ 3.23^{+ 0.32}_{- 1.56}$ & $2.60^{+0.08}_{-0.07}$ & $ -8.32^{+  0.20}_{-  0.15}$ \\
GN13 & 0.48 & $ 0.33^{+ 0.29}_{- 0.07}$ & $ 9.98^{+ 0.04}_{- 0.07}$ & $ 8.42^{+ 0.23}_{- 0.31}$ & $-1.56^{+ 0.25}_{- 0.32}$ & $11.17^{+ 0.04}_{- 0.03}$ & $ 6.87^{+ 7.04}_{- 2.64}$ & $ 1.69^{+ 0.46}_{- 0.26}$ & $1.05^{+0.11}_{-2.32}$ & $ -8.98^{+  0.20}_{-  2.24}$ \\
GN15 & 2.74 & $ 0.21^{+ 0.08}_{- 0.08}$ & $10.97^{+ 0.10}_{- 0.10}$ & $ 8.95^{+ 0.35}_{- 0.31}$ & $-2.00^{+ 0.35}_{- 0.34}$ & $12.37^{+ 0.13}_{- 0.18}$ & $ 3.69^{+ 2.78}_{- 1.28}$ & $ 0.99^{+ 0.20}_{- 0.18}$ & $2.32^{+0.19}_{-0.18}$ & $ -8.62^{+  0.20}_{-  0.30}$ \\
GN20\dag & 4.06 & $ 0.04^{+ 0.02}_{- 0.03}$ & $11.07^{+ 0.04}_{- 0.04}$ & $ 9.56^{+ 0.22}_{- 0.13}$ & $-1.50^{+ 0.30}_{- 0.16}$ & $13.27^{+ 0.02}_{- 0.03}$ & $ 8.66^{+ 0.01}_{- 1.88}$ & $ 1.09^{+ 0.15}_{- 0.38}$ & $3.34^{+0.05}_{-0.02}$ & $ -7.72^{+  0.01}_{-  0.01}$ \\
GN25 & 1.01 & $ 0.39^{+ 0.09}_{- 0.09}$ & $11.15^{+ 0.13}_{- 0.11}$ & $ 8.97^{+ 0.23}_{- 0.25}$ & $-2.19^{+ 0.26}_{- 0.27}$ & $11.97^{+ 0.03}_{- 0.05}$ & $ 6.61^{+ 5.20}_{- 2.44}$ & $ 1.81^{+ 0.18}_{- 0.18}$ & $1.83^{+0.06}_{-0.08}$ & $ -9.32^{+  0.15}_{-  0.20}$ \\
GN26 & 1.22 & $ 0.07^{+ 0.05}_{- 0.04}$ & $10.72^{+ 0.09}_{- 0.20}$ & $ 8.67^{+ 0.19}_{- 0.13}$ & $-2.02^{+ 0.23}_{- 0.18}$ & $12.58^{+ 0.03}_{- 0.03}$ & $ 4.35^{+ 2.36}_{- 1.06}$ & $ 1.09^{+ 0.18}_{- 0.56}$ & $2.64^{+0.05}_{-0.04}$ & $ -8.07^{+  0.25}_{-  0.15}$ \\
GN31 & 0.94 & $ 0.37^{+ 0.48}_{- 0.10}$ & $10.67^{+ 0.13}_{- 0.12}$ & $ 9.01^{+ 0.16}_{- 0.22}$ & $-1.67^{+ 0.22}_{- 0.24}$ & $11.52^{+ 0.06}_{- 0.06}$ & $ 3.67^{+ 2.86}_{- 1.72}$ & $ 0.89^{+ 0.38}_{- 0.20}$ & $1.41^{+0.12}_{-1.63}$ & $ -9.32^{+  0.25}_{-  1.70}$ \\
GN34 & 1.36 & $ 0.09^{+ 0.05}_{- 0.05}$ & $10.05^{+ 0.09}_{- 0.07}$ & $ 7.89^{+ 0.68}_{- 0.43}$ & $-2.17^{+ 0.69}_{- 0.44}$ & $11.74^{+ 0.05}_{- 0.07}$ & $ 2.63^{+ 1.36}_{- 1.18}$ & $ 0.47^{+ 0.16}_{- 0.18}$ & $1.79^{+0.08}_{-0.07}$ & $ -8.23^{+  0.10}_{-  0.20}$ \\
LESS010 & 2.44 & $ 0.05^{+ 0.02}_{- 0.03}$ & $10.34^{+ 0.22}_{- 0.01}$ & $ 9.35^{+ 0.29}_{- 0.20}$ & $-1.04^{+ 0.31}_{- 0.24}$ & $12.63^{+ 0.10}_{- 0.01}$ & $ 3.81^{+ 0.68}_{- 0.01}$ & $ 0.45^{+ 0.34}_{- 0.01}$ & $2.74^{+0.07}_{-0.01}$ & $ -7.62^{+  0.01}_{-  0.15}$ \\
LESS011 & 2.68 & $ 0.05^{+ 0.11}_{- 0.04}$ & $10.61^{+ 0.34}_{- 0.25}$ & $ 9.28^{+ 0.14}_{- 0.13}$ & $-1.31^{+ 0.28}_{- 0.40}$ & $12.59^{+ 0.05}_{- 0.04}$ & $ 4.19^{+ 5.64}_{- 1.04}$ & $ 1.31^{+ 1.24}_{- 0.94}$ & $2.67^{+0.07}_{-0.12}$ & $ -7.93^{+  0.25}_{-  0.45}$ \\
LESS018 & 2.21 & $ 0.04^{+ 0.03}_{- 0.03}$ & $10.73^{+ 0.43}_{- 0.34}$ & $ 9.23^{+ 0.11}_{- 0.11}$ & $-1.48^{+ 0.35}_{- 0.36}$ & $12.90^{+ 0.01}_{- 0.08}$ & $ 2.39^{+ 2.12}_{- 0.22}$ & $ 0.41^{+ 0.50}_{- 0.10}$ & $3.02^{+0.06}_{-0.12}$ & $ -7.82^{+  0.50}_{-  0.30}$ \\
LESS040 & 1.59 & $ 0.05^{+ 0.02}_{- 0.02}$ & $10.01^{+ 0.01}_{- 0.02}$ & $ 8.94^{+ 0.23}_{- 0.18}$ & $-1.06^{+ 0.23}_{- 0.19}$ & $12.09^{+ 0.08}_{- 0.01}$ & $ 5.23^{+ 0.01}_{- 2.20}$ & $ 0.95^{+ 0.01}_{- 0.44}$ & $2.20^{+0.10}_{-0.01}$ & $ -7.82^{+  0.10}_{-  0.01}$ \\
LESS067 & 2.12 & $ 0.05^{+ 0.05}_{- 0.03}$ & $10.60^{+ 0.51}_{- 0.23}$ & $ 8.96^{+ 0.24}_{- 0.19}$ & $-1.64^{+ 0.30}_{- 0.35}$ & $12.65^{+ 0.06}_{- 0.05}$ & $ 2.71^{+ 3.58}_{- 0.56}$ & $ 0.69^{+ 0.28}_{- 0.22}$ & $2.76^{+0.04}_{-0.05}$ & $ -7.82^{+  0.20}_{-  0.55}$ \\
LESS079 & 2.07 & $ 0.05^{+ 0.04}_{- 0.04}$ & $10.70^{+ 0.17}_{- 0.23}$ & $ 9.03^{+ 0.17}_{- 0.14}$ & $-1.65^{+ 0.25}_{- 0.27}$ & $12.66^{+ 0.05}_{- 0.04}$ & $ 2.57^{+ 1.06}_{- 0.72}$ & $ 0.87^{+ 0.10}_{- 0.40}$ & $2.72^{+0.06}_{-0.02}$ & $ -7.93^{+  0.15}_{-  0.25}$ \\
LOCK850.03 & 3.04 & $ 0.08^{+ 0.03}_{- 0.02}$ & $11.23^{+ 0.01}_{- 0.02}$ & $ 9.48^{+ 0.24}_{- 0.26}$ & $-1.75^{+ 0.28}_{- 0.30}$ & $12.94^{+ 0.01}_{- 0.03}$ & $ 1.59^{+ 0.32}_{- 0.01}$ & $ 0.49^{+ 0.80}_{- 0.01}$ & $3.03^{+0.01}_{-0.09}$ & $ -8.18^{+  0.01}_{-  0.05}$ \\
LOCK850.12 & 2.47 & $ 0.14^{+ 0.09}_{- 0.09}$ & $11.22^{+ 0.11}_{- 0.14}$ & $ 9.14^{+ 0.26}_{- 0.22}$ & $-2.06^{+ 0.28}_{- 0.26}$ & $12.72^{+ 0.03}_{- 0.10}$ & $ 6.29^{+ 2.64}_{- 2.04}$ & $ 1.31^{+ 0.14}_{- 0.34}$ & $2.71^{+0.14}_{-0.13}$ & $ -8.48^{+  0.10}_{-  0.15}$ \\
LOCK850.14 & 2.61 & $ 0.22^{+ 0.08}_{- 0.07}$ & $11.27^{+ 0.07}_{- 0.08}$ & $ 9.20^{+ 0.24}_{- 0.21}$ & $-2.06^{+ 0.25}_{- 0.24}$ & $12.66^{+ 0.07}_{- 0.08}$ & $ 4.05^{+ 3.62}_{- 1.60}$ & $ 0.99^{+ 0.18}_{- 0.14}$ & $2.62^{+0.08}_{-0.09}$ & $ -8.62^{+  0.10}_{-  0.15}$ \\
LOCK850.15\dag & 2.76 & $ 0.05^{+ 0.03}_{- 0.03}$ & $10.63^{+ 0.22}_{- 0.02}$ & $ 9.22^{+ 0.33}_{- 0.30}$ & $-1.47^{+ 0.36}_{- 0.33}$ & $12.90^{+ 0.02}_{- 0.10}$ & $ 2.95^{+ 2.04}_{- 0.18}$ & $ 1.07^{+ 0.22}_{- 0.70}$ & $2.95^{+0.08}_{-0.09}$ & $ -7.68^{+  0.01}_{-  0.30}$ \\
LOCK850.16 & 1.62 & $ 0.29^{+ 0.09}_{- 0.13}$ & $11.53^{+ 0.06}_{- 0.11}$ & $ 9.06^{+ 0.31}_{- 0.26}$ & $-2.47^{+ 0.33}_{- 0.28}$ & $12.60^{+ 0.05}_{- 0.05}$ & $ 7.71^{+ 4.36}_{- 4.34}$ & $ 2.09^{+ 0.18}_{- 0.64}$ & $2.52^{+0.07}_{-0.12}$ & $ -9.02^{+  0.20}_{-  0.15}$ \\
SMMJ105238+571651 & 1.85 & $ 0.05^{+ 0.02}_{- 0.03}$ & $10.47^{+ 0.18}_{- 0.33}$ & $ 8.83^{+ 0.54}_{- 0.54}$ & $-1.62^{+ 0.54}_{- 0.56}$ & $12.48^{+ 0.05}_{- 0.01}$ & $ 4.31^{+ 0.24}_{- 2.52}$ & $ 0.51^{+ 0.28}_{- 0.16}$ & $2.62^{+0.04}_{-0.05}$ & $ -7.88^{+  0.40}_{-  0.15}$ \\
\hline                                                                                 
\end{tabular}                                                                                                      
\label{tab:all_SMG_properties}
\end{center}
\end{table*}

\section{{\sc magphys} star-formation histories}
The SFHs derived from our SED fitting in \S\ref{sec:results}, are shown in Fig.~\ref{fig:SFHs}.

\begin{figure*}
\includegraphics[width=17.5cm, height=23.1cm]{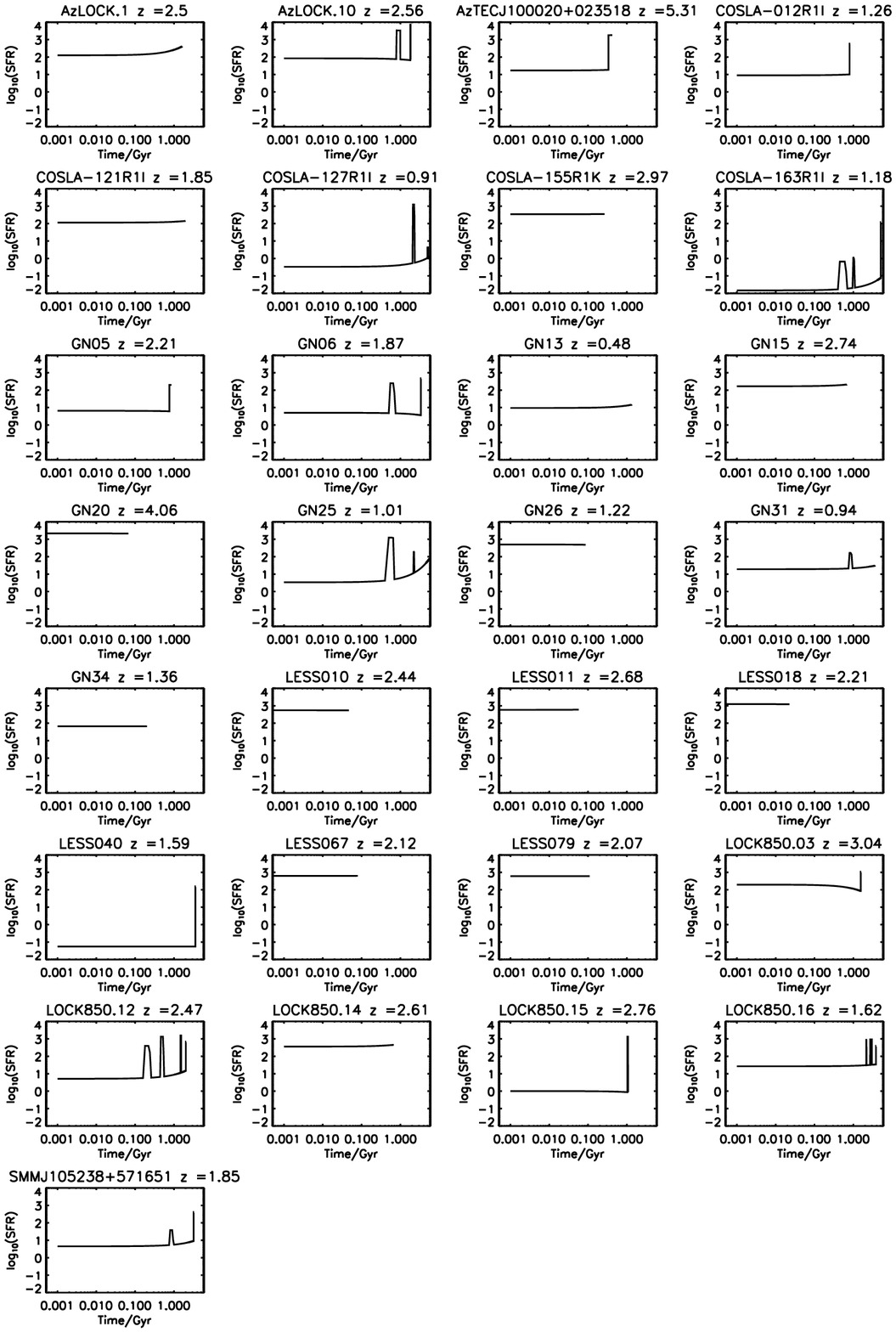}
\caption{Best-fit SFHs of the 29 SMGs with good SED fits derived from {\sc magphys} SED fitting. The majority of SFHs can be described as `bursts' of star formation, either because they have a short elevated SFR near the current age, or because their SFHs are so short and extreme they can be considered a burst.}
\label{fig:SFHs}
\end{figure*}

\label{lastpage}

\end{document}